\documentclass[preprint2]{aastex}
\usepackage{color}
\usepackage{epsfig}
\usepackage[fleqn]{amsmath}
\usepackage{multirow}
\usepackage{mathptmx}
\usepackage{graphicx}
\usepackage{bm}

\shorttitle{Energy-conserving integrator} \shortauthors{Hu et al.}

\begin{document}


\title{Energy-conserving integrator for conservative Hamiltonian systems with ten-dimensional
phase space }


\author{Shiyang Hu$^{1}$, Xin Wu$^{1,2,3, \dag}$, Enwei Liang$^{1,3}$}
\affil{1. School of Physical Science and Technology, Guangxi
University,
Nanning 530004, China \\
2. School of Mathematics, Physics and Statistics $\&$
Center of Application and Research of Computational Physics, Shanghai
University of Engineering Science, Shanghai 201620, China \\
3. Guangxi Key Laboratory for Relativistic Astrophysics, Guangxi
University, Nanning 530004, China} \email{$\dag$ Corresponding
Author wuxin$\_$1134@sina.com (X. W.); 
2312639147@qq.com (S. H.), lew@gxu.edu.cn (E. L.)}



\begin{abstract}

In this paper, an implicit nonsymplectic exact energy-preserving integrator is
specifically designed for a ten-dimensional phase-space
conservative Hamiltonian system with five degrees of freedom. It
is based on  a suitable discretization-averaging of the
Hamiltonian gradient, with a second-order accuracy to numerical
solutions. A one-dimensional disordered discrete nonlinear
Schr\"{o}dinger equation and a post-Newtonian Hamiltonian system of
spinning compact binaries are taken as our two examples. We demonstrate
numerically that the proposed algorithm exhibits good long-term
performance in the preservation of energy, if roundoff errors are
neglected. This result is independent of time steps, initial
orbital eccentricities, and regular and chaotic orbital dynamical
behavior. In particular, the application of appropriately large
time steps to the new algorithm is helpful in reducing
time-consuming and roundoff errors. This new method, combined with
fast Lyapunov indicators, is well suited related to chaos in the two example problems. It is found that chaos in the former system is mainly
responsible for one of the parameters. In the latter problem, a
combination of small initial separations and high initial
eccentricities can easily induce chaos.

\end{abstract}


\emph{Unified Astronomy Thesaurus concepts}: Black hole physics
(159); Computational methods (1965); Computational astronomy
(293); Celestial mechanics (211)



\section{Introduction}

Recently, several gravitational-wave signals (e.g. GW150914 and
GW190521) emitted from binary black hole mergers were successfully
detected (Abbott et al. 2016, 2020a, 2020b). These are consistent
with the predictions of Einstein's theory of general relativity.
The black holes' masses and spins were also inferred from the
observed data. The post-Newtonian (PN) expansion of the
relativistic two-body dynamics (Blanchet et al. 1995; Buonanno
$\&$ Damour 1999; Blanchet 2014) is suitable for describing
theoretical templates of the early inspiral gravitational
waveforms. The known PN equations of motion are accurate to
$O((v/c)^6)$ [i.e., third post-Newtonian (3PN) order], and are even
accurate to $O((v/c)^7)$ (Blanchet 2014), where the orbital
velocity, $v$, is small, as compared with the speed of light, $c$.
Harmonic-coordinate Lagrangian formalism and ADM-coordinate
Hamiltonian formalism were independently applied to provide the PN
equations of motion. On the basis of the Legendre transform, the
former formalism can be transformed into the latter;
inversely, the former formalism can also be derived from the
latter. This being the case, Damour et al. (2001a) and de Andrade et al.
(2001) claimed that the two methods for describing the motion of
two compact bodies are physically equivalent. Higher-order PN
terms are truncated in general when one of  the two PN formalisms
is derived from another. This fact shows that the two PN
formalisms have some differences (Wu et al. 2015; Wu $\&$ Huang
2015). The coherent non-truncated equations of motion derived from
a PN Lagrangian also differ slightly differ from the truncated equations
of motion derived from the PN Lagrangian (Li et al. 2019, 2020).
Specifically, some small differences always exist between the coherent
Lagrangian equations, the truncated Lagrangian equations, and the
Hamiltonian at the same PN order. In some circumstances, such small
differences cause the three PN presentations to have complete
different orbital dynamical behavior in terms of order and chaos.

When the black hole binaries are non-spinning, the PN equations of
motion are analytically integrable and solvable from a theoretical
point of view. However, the analytical solutions of the PN
equations are difficult to be written as explicit functions of
time. If the bodies are spinning, the PN equations of motion
become non-integrable. The binaries may even exhibit chaos for the
description of a dynamical system exhibiting sensitive
dependence to initial conditions. The onset of chaos may affect a matched filtering method requiring theoretical templates
of gravitational waves that agree with experimental gravitational-wave signals, without noise. On the other hand, it can enhance the
signals, in which case it may be helpful to observe the gravitational
waves (Levin 1999; Cornish $\&$ Levin 2002). Considering the fact that
chaos may make either a positive or negative contribution to the detection of
the waveforms, several authors have focused on the chaotic dynamics of
spinning black hole binaries (Schnittman $\&$ Rasio 2001;
K\"{o}nigsd\"{o}rffer $\&$ Gopakumar 2005; Gopakumar $\&$
K\"{o}nigsd\"{o}rffer 2005; Levin 2006; Wu $\&$ Huang 2015) finding chaos in some cases (Levin 2000; Levin 2003; Cornish $\&$
Levin 2003; Hartl $\&$ Buonanno 2005; Wu $\&$ Xie 2007, 2008; Wang
$\&$ Wu 2011; Huang et al. 2014; Wu et al. 2015; Huang et al.
2016). In the presence of chaos, analytically solving the PN equations of
motion for the systems of spinning black hole binaries
analytically is impossible.

Analytically solving the PN equations of motion presents
considerable difficulties, regardless of whether the binaries are spinning or not. However,
numerically solving the PN equations is very easy and convenient.
Conventional integrators, such as Runge-Kutta methods, often lead
to an increase in energy errors with time, and very poor energy
accuracy in long integration time, so that the obtained results are
unreliable. At this point, a note is worthwhile on the subject of energy
conservation. In fact, energy conservation is essential in
numerical integration. It is a fundamental property inherent in
conservative Hamiltonian flow. In addition, it is often
used to check numerical accuracy, although it is not always
completely reliable in obtaining high-precision results. Enforcing
the conservation of energy significantly improves the quality of
orbit integrations in many situations. Below, we
introduce three paths for energy conservation.

Smplectic integrators (Wisdom 1982; Ruth 1983) preserve
the original Hamiltonian flow. They do not exactly conserve
the real Hamiltonian or energy, but they cause the energy to
oscillate. Specifically, they show no secular energy drift,
causing the energy errors to remain bounded. As such,
symplectic integrators are regarded to be energy-conserving.
Due to these good properties, symplectic integrators, such as
the second-order symplectic method of Wisdom $\&$ Holman
(1991), are widely used in the long-term integration of celestial
N-body problems in the solar system. Explicit symplectic
methods are less expensive than implicit methods, and
therefore have priority in terms of application. However, they
are not available to relativistic or PN problems in general, as
either the variables in the systems are inseparable, or two
integrable splitting parts can be given to the systems, but their
analytical solutions are not explicit functions of time. As a
result, implicit symplectic algorithms, such as the implicit
midpoint rule (Feng 1986), are considered. Explicit and
implicit symplectic composition methods (Liao 1997; Preto
$\&$ Saha 2009; Lubich et al. 2010; Mei et al. 2013a) are superior
to completely implicit symplectic integrators with respect to
computational efficiency. Such a construction was encountered
in the simulation of the PN Hamiltonian dynamics of spinning
compact binaries (Zhong et al. 2010; Mei et al. 2013b). More
recently, explicit symplectic integrators were proposed for
Schwarzschild spacetime geometry, whose Hamiltonian can be
split into four integrable parts, with analytical solutions as
explicit functions of proper time (Wang et al. 2021a). They are
also applicable to the Hamiltonian of Reissner-Nordstr\"{o}m
black holes (Wang et al. 2021b). On the other hand, the
extended phase-space explicit methods employed by Pihajoki
(2015) in relation to inseparable Hamiltonian problems possess
energy conservation, similarly to symplectic methods. Permutations of coordinates and/or momenta have been optimized by
Liu et al. (2016) and Luo et al. (2017).

The manifold correction scheme of Nacozy (1971) pulls the
numerical solution (given by a conventional integrator) back to
the original manifold of constant energy along the least-squares
shortest path. The corrected solution approximately satisfies the
constant energy; it can cause the energy to be accurate to a
machine double-precision if the integrator provides a machine
single-precision to the energy. Using integral invariant
relations, this approach has been extended to adjust the
integrated position and velocity so as to approximately satisfy
the varying Kepler energy, Laplace integral, and/or angular
momentum vector for each of the N bodies in the solar system
(Wu et al. 2007; Ma et al. 2008a). The least-squares correction
approach has also proven applicable to the numerical correction
of all integrals in the conservative PN Hamiltonian formulation
of spinning compact binaries (Zhong $\&$ Wu 2010). In terms of
scale transformations, Nacozy's idea was also further developed to rigorously satisfy the varying Kepler energy, Laplace
integral, and/or angular momentum vector at every integration
step (Fukushima 2003a, 2003b, 2003c, 2004; Ma et al. 2008b).
The velocity-scaling correction method has been applied to
nonconservative or dissipative restricted three-body problems
(Wang et al. 2016, 2018). Recently, the Kepler solver was used
to rigorously conserve all integrals and orbital elements (with
the exception of the mean longitude) of quasi-Keplerian orbits
(Deng et al. 2020). These correction schemes have been shown
to significantly suppress the growth of integration errors, and to
dramatically enhance the quality of the integrations.

In addition to a variety of correction schemes, a class of
energy-preserving numerical integration methods can exactly
preserve the energies of Hamiltonian systems. These are
obtained by suitably discretizing the canonical equations of
Hamiltonian systems. Chorin et al. (1978) constructed an
energy-conserving algorithm for the discretization of Hamiltonian gradients with a first-order accuracy to the numerical
solutions. A similar integrator was also proposed by Feng
(1985). A second-order approximation to the Hamiltonian
gradient has been designed for a four-dimensional phase-space
Hamiltonian system (Qin 1987; Itoh $\&$ Abe 1988). Here, each
component of the Hamiltonian gradient is an average of four
difference terms, each of which is the ratio of the increment of
the Hamiltonian between positions (or momenta) to the
position (or momentum) increment. Such discretization-averaging of the Hamiltonian equations is complex, because it
depends on the dimension of a Hamiltonian phase space.
Bacchini et al. (2018, 2019) established a precise energy-conserving implicit integration scheme for six-dimensional
phase-space Hamiltonian problems, and simulated time-like
(massive particles) and null (photons) geodesics in Schwarzschild and Kerr spacetimes. This idea was then generalized to
the eight-dimensional phase-space PN Hamiltonian dynamics
of compact binaries with one body spinning (Hu et al. 2019).
These energy-conserving schemes are based on the coordinate-increment discrete gradient (Robert et al. 1999). They can also
be constructed in terms of the mean value discrete gradient
(Harten 1983; Quispel $\&$ McLaren 2008; Huang $\&$ Mei 2020)
and the midpoint discrete gradient (Gonzalez 1996).

Following on from our previous work (Hu et al. 2019), we
introduce a new energy-conserving algorithm for a ten-dimensional phase-space PN Hamiltonian of black hole
binaries. This scheme gives second-order approximations to
numerical solutions. This is the main aim of this paper. The
remainder of this paper is organized as follows: in Section 2,
we describe how to construct an energy-conserving method for
ten-dimensional phase-space Hamiltonian problems. Taking a
one-dimensional disordered discrete nonlinear Schr\"{o}dinger
equation (Skokos et al. 2014; Senyange et al. 2018) as an
example, we evaluate the numerical performance of the new
algorithm. The new method is applied to study the dynamical
properties of the system in Section 3. For comparison, the
Runge-Kutta method, the implicit midpoint method
(Feng 1986; Zhong et al. 2010; Mei et al. 2013a), and an
extended phase-space symplectic-like algorithm (Pihajoki 2015;
Liu et al. 2016; Li $\&$ Wu 2017; Luo et al. 2017) are also used.
In Section 4, we choose a PN Hamiltonian of spinning compact
binaries as another example with which to verify the
algorithmic performance. The new method is used to investigate the regular and chaotic dynamical behavior of orbits
in the PN problem. Section 5 summarises our main results and
conclusions. The new method is shown to obtain the numerical
solutions, accurate to the second order, in the Appendix.

\section{Construction of new energy-conserving method}

Let $\bm{q}$ and $\bm{p}$, respectively, be $N$-dimensional generalized coordinate and conjugate momentum
vectors in a $2N$-dimensional conservative Hamiltonian system, $H(\bm{q},\bm{p})$.
This Hamiltonian corresponds to the following canonical equations:
\begin{equation}\label{1}
  \dot{\bm{q}} = \frac{\partial {\emph{H}}}{\partial {\bm{p}}},
\end{equation}
\begin{equation}\label{2}
  \dot{\bm{p}} = -\frac{\partial {\emph{H}}}{\partial {\bm{q}}}.
\end{equation}
Taking time step $h=t_{n+1}-t_n$ and discretizing the canonical
equations, we obtain  the relationship between the solution
$(p_n,q_n)$ at the $n$th step, and the solution $(p_{n+1},q_{n+1})$
at the $(n+1)$th step as follows:
\begin{equation}\label{3}
  \frac{\bm{q}_{n+1}-\bm{q}_{n}}{h} = \frac{H(\bm{q}_{n},\bm{p}_{n+1})-H(\bm{q}_{n},\bm{p}_{n})}{\bm{p}_{n+1}-\bm{p}_{n}},
\end{equation}
\begin{equation}\label{4}
  \frac{\bm{p}_{n+1}-\bm{p}_{n}}{h} =- \frac{H(\bm{q}_{n+1},\bm{p}_{n+1})-H(\bm{q}_{n},\bm{p}_{n+1})}{\bm{q}_{n+1}-\bm{q}_{n}}.
\end{equation}
It can be derived from Equations \eqref{3} and \eqref{4} that
\begin{equation}\label{5}
H(\bm{q}_{n+1},\bm{p}_{n+1})=H(\bm{q}_{n},\bm{p}_{n}).
\end{equation}
This result shows that Equations \eqref{3} and \eqref{4} are an exact energy-conserving algorithm,
which provides the solution $(p_{n+1},q_{n+1})$ at the $(n+1)$th step, after the solution $(p_n,q_n)$ at the $n$th step
advances the time step, $h$. In general, the solution is implicitly given. In this case, the use of a certain iteration method is necessary.

In fact, the above energy-conserving method belongs to a
Hamiltonian-conserving method\footnote{Generally, a Hamiltonian
corresponds to the energy of a system. In this case, the
Hamiltonian-conserving means the energy-conserving. However, a
Hamiltonian is not always equivalent to the energy of a system in
some cases. For example, the Hamiltonian for a circular restricted
three-body Hamiltonian problem in an inertial frame explicitly
depends on time, and therefore is not the Jacobian constant in the
inertial frame (Su et al. 2016). Such a Hamiltonian-conserving
does not mean an energy-conserving.}, which is only viewed as a
first-order approximation to the Hamiltonian gradient. It therefore
provides a first-order accuracy to the numerical solutions. In
fact, there is also a second-order Hamiltonian-conserving scheme
for four-dimensional phase-space Hamiltonian systems (Qin 1987;
Itoh $\&$ Abe 1988; Feng $\&$ Qin 2009), as discussed in the
introduction. In that algorithmic construction, a second-order
discrete approximation to the Hamiltonian gradient is an average
of four Hamiltonian difference terms. The average becomes more
complicated as the dimensionality of the Hamiltonian increases.

Following on from previous work of Hu et al. (2019), we consider the
construction of a Hamiltonian-conserving scheme for a ten-dimensional
phase-space conservative system. For simplicity, 0 and 1
respectively correspond to the numbers of integration steps, $n$
and $(n+1)$; e.g. $H(q_{1(n+1)}$, $q_{2(n+1)}$, $q_{3(n+1)}$,
$q_{4(n+1)}$, $q_{5(n+1)}$, $p_{1(n+1)}$, $p_{2(n+1)}$,
$p_{3(n+1)}$, $p_{4(n+1)}$, $p_{5(n+1)})/$ $(p_{1(n+1)}-p_{1(n)})$
= $H(1111111111)/(p_{11}-p_{10})$. Equations \eqref{1} and
\eqref{2} are discretized in the following forms:
\begin{eqnarray}\label{6}
&& \frac{q_{11}-q_{10}}{h} = \frac{1}{10} \frac{1}{p_{11}-p_{10}} \nonumber \\
&& \times \left[H(0000010000)-H(0000000000)\right. \nonumber \\
&& +H(10000 10000)-H(10000 00000) \nonumber \\
&& +H(00001 10001)-H(00001 00001) \nonumber \\
&& +H(11000 11000)-H(11000 01000) \nonumber \\
&& +H(00011 10011)-H(00011 00011) \nonumber \\
&& +H(11100 11100)-H(11100 01100) \nonumber \\
&& +H(00111 10111)-H(00111 00111) \nonumber \\
&& +H(11110 11110)-H(11110 01110) \nonumber \\
&& +H(01111 11111)-H(01111 01111) \nonumber \\
&& \left.+H(1111111111)-H(1111101111) \right],
\end{eqnarray}
\begin{eqnarray}\label{7}
&& \frac{q_{21}-q_{20}}{h} = \frac{1}{10} \frac{1}{p_{21}-p_{20}} \nonumber \\
&& \times \left[H(0000001000)-H(0000000000)\right. \nonumber \\
&& +H(01000 01000)-H(01000 00000) \nonumber \\
&& +H(10000 11000)-H(10000 10000) \nonumber \\
&& +H(01100 01100)-H(01100 00100) \nonumber \\
&& +H(10001 11001)-H(10001 10001) \nonumber \\
&& +H(01110 01110)-H(01110 00110) \nonumber \\
&& +H(10011 11011)-H(10011 10011) \nonumber \\
&& +H(01111 01111)-H(01111 00111) \nonumber \\
&& +H(10111 11111)-H(10111 10111) \nonumber \\
&& \left.+H(1111111111)-H(1111110111) \right],
\end{eqnarray}
\begin{eqnarray}\label{8}
&& \frac{q_{31}-q_{30}}{h} = \frac{1}{10} \frac{1}{p_{31}-p_{30}} \nonumber \\
&& \times \left[H(0000000100)-H(0000000000)\right. \nonumber \\
&& +H(00100 00100)-H(00100 00000) \nonumber \\
&& +H(01000 01100)-H(01000 01000) \nonumber \\
&& +H(00110 00110)-H(00110 00010) \nonumber \\
&& +H(11000 11100)-H(11000 11000) \nonumber \\
&& +H(00111 00111)-H(00111 00011) \nonumber \\
&& +H(11001 11101)-H(11001 11001) \nonumber \\
&& +H(10111 10111)-H(10111 10011) \nonumber \\
&& +H(11011 11111)-H(11011 11011) \nonumber \\
&& \left.+H(1111111111)-H(1111111011) \right],
\end{eqnarray}
\begin{eqnarray}\label{9}
&& \frac{q_{41}-q_{40}}{h} = \frac{1}{10} \frac{1}{p_{41}-p_{40}} \nonumber \\
&& \times \left[H(0000000010)-H(0000000000)\right. \nonumber \\
&& +H(00010 00010)-H(00010 00000) \nonumber \\
&& +H(00100 00110)-H(00100 00100) \nonumber \\
&& +H(00011 00011)-H(00011 00001) \nonumber \\
&& +H(01100 01110)-H(01100 01100) \nonumber \\
&& +H(10011 10011)-H(10011 10001) \nonumber \\
&& +H(11100 11110)-H(11100 11100) \nonumber \\
&& +H(11011 11011)-H(11011 11001) \nonumber \\
&& +H(11101 11111)-H(11101 11101) \nonumber \\
&& \left.+H(1111111111)-H(1111111101) \right],
\end{eqnarray}
\begin{eqnarray}\label{10}
&& \frac{q_{51}-q_{50}}{h} = \frac{1}{10} \frac{1}{p_{51}-p_{50}} \nonumber \\
&& \times \left[H(0000000001)-H(0000000000)\right. \nonumber \\
&& +H(00001 00001)-H(00001 00000) \nonumber \\
&& +H(00010 00011)-H(00010 00010) \nonumber \\
&& +H(10001 10001)-H(10001 10000) \nonumber \\
&& +H(00110 00111)-H(00110 00110) \nonumber \\
&& +H(11001 11001)-H(11001 11000) \nonumber \\
&& +H(01110 01111)-H(01110 01110) \nonumber \\
&& +H(11101 11101)-H(11101 11100) \nonumber \\
&& +H(11110 11111)-H(11110 11110) \nonumber \\
&& \left.+H(1111111111)-H(1111111110) \right],
\end{eqnarray}
\begin{eqnarray}\label{11}
&& \frac{p_{11}-p_{10}}{h} = -\frac{1}{10} \frac{1}{q_{11}-q_{10}} \nonumber \\
&& \times \left[H(1000000000)-H(0000000000)\right. \nonumber \\
&& +H(10000 10000)-H(00000 10000) \nonumber \\
&& +H(11000 01000)-H(01000 01000) \nonumber \\
&& +H(10001 10001)-H(00001 10001) \nonumber \\
&& +H(11100 01100)-H(01100 01100) \nonumber \\
&& +H(10011 10011)-H(00011 10011) \nonumber \\
&& +H(11110 01110)-H(01110 01110) \nonumber \\
&& +H(10111 10111)-H(00111 10111) \nonumber \\
&& +H(11111 01111)-H(01111 01111) \nonumber \\
&& \left.+H(1111111111)-H(0111111111) \right],
\end{eqnarray}
\begin{eqnarray}\label{12}
&& \frac{p_{21}-p_{20}}{h} = -\frac{1}{10} \frac{1}{q_{21}-q_{20}} \nonumber \\
&& \times \left[H(0100000000)-H(0000000000)\right. \nonumber \\
&& +H(01000 01000)-H(00000 01000) \nonumber \\
&& +H(01100 00100)-H(00100 00100) \nonumber \\
&& +H(11000 11000)-H(10000 11000) \nonumber \\
&& +H(01110 00110)-H(00110 00110) \nonumber \\
&& +H(11001 11001)-H(10001 11001) \nonumber \\
&& +H(01111 00111)-H(00111 00111) \nonumber \\
&& +H(11011 11011)-H(10011 11011) \nonumber \\
&& +H(11111 10111)-H(10111 10111) \nonumber \\
&& \left.+H(1111111111)-H(1011111111) \right],
\end{eqnarray}
\begin{eqnarray}\label{13}
&& \frac{p_{31}-p_{30}}{h} = -\frac{1}{10} \frac{1}{q_{31}-q_{30}} \nonumber \\
&& \times \left[H(0010000000)-H(0000000000)\right. \nonumber \\
&& +H(00100 00100)-H(00000 00100) \nonumber \\
&& +H(00110 00010)-H(00010 00010) \nonumber \\
&& +H(01100 01100)-H(01000 01100) \nonumber \\
&& +H(00111 00011)-H(00011 00011) \nonumber \\
&& +H(11100 11100)-H(11000 11100) \nonumber \\
&& +H(10111 10011)-H(10011 10011) \nonumber \\
&& +H(11101 11101)-H(11001 11101) \nonumber \\
&& +H(11111 11011)-H(11011 11011) \nonumber \\
&& \left.+H(1111111111)-H(1101111111) \right],
\end{eqnarray}
\begin{eqnarray}\label{14}
&& \frac{p_{41}-p_{40}}{h} = -\frac{1}{10} \frac{1}{q_{41}-q_{40}} \nonumber \\
&& \times \left[H(0001000000)-H(0000000000)\right. \nonumber \\
&& +H(00010 00010)-H(00000 00010) \nonumber \\
&& +H(00011 00001)-H(00001 00001) \nonumber \\
&& +H(00110 00110)-H(00100 00110) \nonumber \\
&& +H(10011 10001)-H(10001 10001) \nonumber \\
&& +H(01110 01110)-H(01100 01110) \nonumber \\
&& +H(11011 11001)-H(11001 11001) \nonumber \\
&& +H(11110 11110)-H(11100 11110) \nonumber \\
&& +H(11111 11101)-H(11101 11101) \nonumber \\
&& \left.+H(1111111111)-H(1110111111) \right],
\end{eqnarray}
\begin{eqnarray}\label{15}
&& \frac{p_{51}-p_{50}}{h} = -\frac{1}{10} \frac{1}{q_{51}-q_{50}} \nonumber \\
&& \times \left[H(0000100000)-H(0000000000)\right. \nonumber \\
&& +H(00001 00001)-H(00000 00001) \nonumber \\
&& +H(10001 10000)-H(10000 10000) \nonumber \\
&& +H(00011 00011)-H(00010 00011) \nonumber \\
&& +H(11001 11000)-H(11000 11000) \nonumber \\
&& +H(00111 00111)-H(00110 00111) \nonumber \\
&& +H(11101 11100)-H(11100 11100) \nonumber \\
&& +H(01111 01111)-H(01110 01111) \nonumber \\
&& +H(11111 11110)-H(11110 11110) \nonumber \\
&& \left.+H(1111111111)-H(1111011111) \right].
\end{eqnarray}
On the right-hand sides of Equations \eqref{6}-\eqref{15}, each
component of the Hamiltonian gradient is approximately expressed
in terms of an average of 10  Hamiltonian differences. Using
\eqref{6}-\eqref{15}, we simply derive the relation
\begin{equation}\label{16}
H(1111111111)-H(0000000000)=0.
\end{equation}
This fact indicates that Equations \eqref{6}-\eqref{15} strictly
preserve the Hamiltonian energy from the theoretical viewpoint.
This construction is a new energy-conserving algorithm, with a
coordinate-increment discrete gradient. In fact, this method gives
a second-order accuracy to the numerical solutions. Itoh $\&$ Abe
(1988) provided a simple explanation as to why a similar energy-conserving method
for a four-dimensional phase-space Hamiltonian problem is accurate to
the order of $h^2$. We also provide a detailed explanation for
Equations \eqref{6}-\eqref{15}, with the numerical solutions
accurate to this order, in the Appendix.

Below, the efficiency of the new energy-conserving
algorithm is verified, base on simulations of two dynamical models.

\section{One-dimensional disordered discrete nonlinear Schr\"{o}dinger equation}

A one-dimensional disordered discrete nonlinear Schr\"{o}dinger
equation (DDNLS) is considered in Section 3.1. The
performance of the newly proposed energy-conserving algorithm is then
checked in Section 3.2. We use our new method, combined with the
technique of fast Lyapunov indicators (Froeschl\'{e} et al. 1997;
Froeschl\'{e} $\&$ Lega 2000; Wu et al. 2006) to provide an insight
into the influences of some dynamical parameters on orbital
dynamics in Section 3.3.

\subsection{Model}

DDNLS describes the  motion of  coupled, nonlinear oscillators in
a crystal lattice.  The so-called ``one-dimension" in the model
refers to the geometric dimensionality of the model being 1, i.e., the dynamical evolution in the nonlinear partial differential
Schr\"{o}dinger equation with respect to one position variable.
However, the canonical variables of the model can be expanded to
any dimensions in a Hamiltonian from the discretized Schr\"{o}dinger equation. DDNLS corresponds to the following dimensionless
Hamiltonian (Skokos et al. 2014; Senyange et al. 2018)
\begin{eqnarray}\label{17}
H(\bm{q}, \bm{p}) &=& \sum_{m=1}^{N} [\frac{\varepsilon_{m}}{2}\left(q_{m}^{2}+p_{m}^{2}\right)
+\frac{\beta}{8}\left(q_{m}^{2}+p_{m}^{2}\right)^{2} \nonumber \\
&& -\left(p_{m} p_{m+1}+q_{m} q_{m+1}\right)].
\end{eqnarray}
Here, $\varepsilon_{m}$ are random on-site energies  in the interval $[-W/2,W/2]$, where $W$ represents a disorder strength of the system,
and $\beta\geq0$ is a nonlinearity strength.
Taking $N=5$, and boundary conditions $q_{0}=p_{0}=q_{N+1}=p_{N+1}=0$, we rewrite the Hamiltonian as
\begin{eqnarray}\label{18}
&& H(q_{1},q_{2},q_{3},q_{4},q_{5},p_{1},p_{2},p_{3},p_{4},p_{5}) \nonumber \\
&& = \frac{\varepsilon_{1}}{2}{}(q_{1}^{2}+p_{1}^{2})+\frac{\beta}{8}{}(q_{1}^{2}+p_{1}^{2})^{2}-(p_{1}p_{2}+q_{1}q_{2}) \nonumber \\
&& + \frac{\varepsilon_{2}}{2}{}(q_{2}^{2}+p_{2}^{2})+\frac{\beta}{8}{}(q_{2}^{2}+p_{2}^{2})^{2}-(p_{2}p_{3}+q_{2}q_{3}) \nonumber \\
&& + \frac{\varepsilon_{3}}{2}{}(q_{3}^{2}+p_{3}^{2})+\frac{\beta}{8}{}(q_{3}^{2}+p_{3}^{2})^{2}-(p_{3}p_{4}+q_{3}q_{4}) \nonumber \\
&& + \frac{\varepsilon_{4}}{2}{}(q_{4}^{2}+p_{4}^{2})+\frac{\beta}{8}{}(q_{4}^{2}+p_{4}^{2})^{2}-(p_{4}p_{5}+q_{4}q_{5}) \nonumber \\
&& + \frac{\varepsilon_{5}}{2}{}(q_{5}^{2}+p_{5}^{2})+\frac{\beta}{8}{}(q_{5}^{2}+p_{5}^{2})^{2}.
\end{eqnarray}
In addition to the Hamiltonian as an integral of energy, the norm
\begin{eqnarray}\label{19}
   Q = \sum_{m=1}^{5} \frac{1}{2}\left(q_{m}^{2}+p_{m}^{2}\right).
\end{eqnarray}
remains invariant (Skokos et al. 2014).

\subsection{Numerical evaluations}

When Equations \eqref{6}-\eqref{15} are applied to the system
\eqref{18}, we call this algorithm EC. For comparison, the
same-order Runge-Kutta method (RK) in Equations (A6) and (A7),
the implicit midpoint method (IS) in Equations (A10) and (A11), and the
extended phase-space symplectic-like algorithm (ES) (Pihajoki
2015; Liu et al. 2016; Li $\&$ Wu 2017; Luo et al. 2017) are
independently used to integrate the system. The time step is given
by $h=0.01$. Orbits 1 and 2 have the same initial conditions,
$q_{i}=(6-i)/10$ and $p_{i}=0$. Parameters $\varepsilon_{m}=0$
are the same. However, $\beta = 1$ for Orbit 1, and $\beta = 10$
for Orbit 2.

Figures 1 (a) and (b)\footnote{The related codes for the figures are
written in Fortran 90, and can be accessed online at
doi.org/10.5281/zenodo.4528966.} plot the Hamiltonian errors for
the four algorithms solving Orbits 1 and 2. When the integration
time reaches $t=10^7$ corresponding to $10^9$ steps, the
Hamiltonian errors calculated by the RK method grow linearly with
time for the two orbits. They are larger than those given by the
IS or ES method. The errors remain bounded and stable for IS.
However, ES gives a secular drift to the energy errors after
$t=10^4$. This may be due to the use of a slightly larger step size. In
fact, this drift is absent if $h=0.001$. Clearly, the EC method
shows the smallest errors with slight secular growths, as compared
with the other three algorithms. The errors for EC are approximate
to those for an eighth- and ninth-order Runge--Kutta--Fehlberg
integrator [RKF89] with adaptive step sizes. Without doubt, the
Hamiltonian or energy is conserved by the IS method because IS is
symplectic (Hairer et al. 2006), and retains this property over a
long-term integration (Rein et al. 2019; Hernandez et al. 2020).
Although the RK and EC methods give secular growths to the
Hamiltonian errors, they are not the same. The RK method does not
preserve the energy, while the EC method does. This is because the
largest errors for the RK method are predominantly
algorithmic truncation errors, and the smallest errors for the EC
method are due to the roundoff errors. The slopes  for the error
growth with time from small to large correspond to algorithms IS,
RKF89, EC, ES, and RK. However, none of the algorithms has zero
slope  for the error growth of the norm in Figures 1 (c) and (d).
The error of the norm is the smallest for RKF89, whereas it is the
largest for RK. In terms of the accuracy of the norm, IS is slightly larger
than RKF89, but smaller than EC. The slopes of the growth of norm
errors with time are $0.6 \sim 0.7$ for IS, and are due to
roundoff errors, due to the symplecticity of IS. However, the
slopes of the growth of norm errors with time for EC are $0.7 \sim
1$, and are mainly due to truncation errors, because EC does not
conserve the norm. The norm errors for other methods (except for
RKF89) are also due to truncation errors.

Owing to its use of many iterations, EC naturally has the
poorest efficiency, as shown in Figure 2. Some details with
regard to drawing the efficiency plot are taken from Rein $\&$
Tamayo (2015) or Deng et al. (2020). RK exhibits the best
efficiency. The efficiency of ES is better than that of IS. In
particular, EC shows the best energy accuracies, which do not
seem to depend on the choice of step sizes. However, the
energy accuracies depend on the choice of step sizes for IS, ES,
and RK. This implies that EC can use a larger step size in order
to reduce computational cost.

The aforementioned numerical tests have confirmed that the
new EC method offers long-term conservation of energy if no
roundoff errors are considered. This result is independent of the
dynamical behavior of the orbits. In fact, Orbit 1 is regular, but
Orbit 2 is chaotic, as described by the techniques of power
spectra and fast Lyapunov indicators (FLIs) in Figure 3. We
make two points relating to the two techniques: the method of
power spectra depicts a distribution of frequencies of orbits. In
general, a regular orbit has discrete spectra, whereas a chaotic
orbit exhibits continuous spectra (Wang $\&$ Wu 2011; Mei et al.
2013b). Using the two different spectra, we can roughly
identify the regularity of Orbit 1, and the chaoticity of Orbit 2,
which are consistently supported by the FLIs obtained from EC
and RKF89 in Figures 3(a)-(d). Although the distinction
between the ordered and chaotic cases can be clearly observed,
the word ``roughly" is still used, because complicated periodic
orbits, quasi-periodic orbits, and weakly chaotic orbits may
have similar continuous spectra, which may also be true of
some non-periodic but non-chaotic orbits. In this sense, other
methods finding chaos, such as FLIs, are necessarily employed. Wu et al. (2006) defined the FLI as
\begin{equation}\label{28}
\emph{FLI}=\log _{10} \frac{d(t)}{d(0)},
\end{equation}
where d(0) and d(t) are the phase-space distances between two
adjacent orbits at times 0 and t, respectively. Clearly, this
describes the growth of the phase-space distance between two
adjacent orbits with time $\log_{10}t$. In fact, it originates from a
modified version of the FLIS of Froeschl\'{e} et al. (1997) and
Froeschl\'{e} $\&$ Lega (2000), as well as a modified version of the
Lyapunov exponents (Tancredi et al. 2001; Wu $\&$ Huang
2003). However, it is more convenient to use than the FLIS of
Froeschl\'{e} $\&$ Lega (2000), and is more sensitive in distinguishing between chaotic and regular bounded orbits than the
technique based on Lyapunov exponents. It grows algebraically
with time for the regular case, but exponentially in the chaotic
case. The completely different time rates for the growth of the
phase-space distance between two adjacent orbits can be used
to distinguish chaos from order. The FLIs in Figures 3(e) and
(f) clearly determine the properties of Orbits 1 and 2.

In short, the new EC method exhibits  good performance in a
long-term numerical integration. It can provide reliable numerical
results, as RKF89 can. Therefore, it is employed to investigate
the orbital dynamics of the DDNLS system.

\subsection{Dependence of chaos on parameters}

Next, let us trace the dynamical transition from order to chaos
with a variation of the parameter $\beta$, or
$\varepsilon_{m}=\varepsilon$. The above initial conditions are
fixed, and the values of $\varepsilon$ are also given in several
values, $\pm0.5$ and $\pm5$. However, $\beta$ runs from 0.01 to 15,
with a span of $\Delta\beta=0.1499$. The FLI is obtained for each
value of $\beta$ after the integration time reaches 3800. It is
found that a value of 5 represents the threshold of FLIs between the ordered and
chaotic cases. FLI$>$5 corresponds to the presence of chaos, and
FLI$\leq$5 indicates the existence of order. Figure 4 plots the
dependence of FLI on $\beta$. Chaos is absent for $\beta<2.4084$.
However, it is present for $\beta>2.4084$, and the intensity of the
chaos increases with an increase in $\beta$. These results are
consistent with those given in Figure 5, which depict the
dependence of FLI on $\varepsilon$ with fixed values of $\beta$.
This is also clearly shown in terms of the FLIs in a two-dimensional
space for parameters $\varepsilon$ and $\beta$ in Figure 6, too.
Regardless of whether $\varepsilon$ is large or small, no chaos
exists for $\beta=1$, whereas chaos exists for $\beta=5,10$,and $15$.
We find similar results when other initial conditions are
considered. We offer the following simple explanation of the result with respect to the
dependence of dynamical transition on the parameters $\beta$ or
$\varepsilon$: if $\beta=0$, the system (18) has only quadratic
terms, and represents a five-dimensional coupled oscillator. In
this case, it is integrable and non-chaotic. This may explain why
the presence or extent of chaos does not depend on $\varepsilon$.
When $\beta\neq0$, the quartic terms cause the system (18) to be
non-integrable, and probably chaotic. If $\beta$ is very small, the
quadratic terms dominate the system (18), and chaos unquestionably does not occur. However, the quartic terms become more important
than the quadratic terms if  $\beta$ is sufficiently large. Only where the
quartic terms approximately match with the quadratic terms, does
chaos become possible. This further explains the data in
Figures 4-6.

Several conclusions can be drawn from the numerical simulations:
$\beta$ is a key parameter for inducing chaos in the DDNLS system.
It has a critical value. Chaos cannot occur when $\beta$
is smaller than this critical value. Furthermore, chaos
has nothing to do with the choice of parameter $\varepsilon_{m}$.

\section{PN Hamiltonian of spinning compact binaries}

In this section, the new EC method is applied to a PN
Hamiltonian of spinning compact binaries, in order to verify
whether this algorithm is still efficient. As such, the
Hamiltonian is introduced in Section 4.1. Numerical estimations are given in Section 4.2. Based on the application of the
EC method, the dynamics of spinning compact binaries are
surveyed in Section 4.3.

\subsection{PN Hamiltonian formulation}

Let two black holes have masses $m_{1}$ and $m_{2}$. The total
mass is $M=m_{1}+m_{2}$. Take the mass ratio $\gamma=m_{1}/ m_{2}$
and the reduced mass $\mu=m_{1}m_{2}/M$. Here,
$\eta=\mu/M=\gamma/(1+\gamma)^{2}$ is a dimensionless parameter;
$\bm{r}=(x,y,z)$ is a position vector of the body, $m_{1}$,
relative to the body, $m_{2}$, and $\bm{n}= \bm{r}/r$ is a
unit radial vector. The two bodies have spins, described by
$\bm{S}_{1}$ and $\bm{S}_{2}$. The speed of light, $c$, and
the gravitational constant, $G$, use geometric units, $c=G=1$.
Dimensionless operations are carried out via scale transformations
as follows: $\bm{r}\rightarrow M \bm{r}$, $t\rightarrow
Mt$, $\bm{S}_i\rightarrow M \mu \bm{S}_i$ $(i=1,2)$ and
$H\rightarrow \mu H$. In addition, $\bm{p}\rightarrow \mu
\bm{p}$ and $\bm{L}\rightarrow M \mu\bm{L}$, where
$\bm{p}$ is a momentum of the body, $m_{1}$, relative to the
body $m_{2}$, and $\bm{L}=\bm{r} \times \bm{p}$ is a
Newtonian-like angular momentum vector. The evolution of
$(\bm{r},\bm{p})$ is governed by the following
dimensionless PN Hamiltonian formulation (Damour et al. 2000a,
2000b, 2001b; Nagar 2011):
\begin{eqnarray}\label{29}
 H(\textbf{r},\textbf{p},\textbf{S}_{1},\textbf{S}_{2}) &=& H_{O}(\textbf{r},\textbf{p})
 + H_{SO}(\textbf{r},\textbf{p},\textbf{S}_{1},\textbf{S}_{2})\nonumber \\
 && + H_{SS}(\textbf{r},\textbf{S}_{1},\textbf{S}_{2}).
\end{eqnarray}
Here, $H_{O}$ is an orbital component, including the Newtonian term and
PN terms to the second order. It is written as
\begin{equation}\label{30}
H_{O} =H_{N}(\textbf{r},\textbf{p})+H_{1PN}(\textbf{r},\textbf{p})+H_{2PN}(\textbf{r},\textbf{p}),
\end{equation}
where the three sub-Hamiltonians are
\begin{equation}\label{31}
H_{N} = \frac{\textbf{p}^2}{2}-\frac{1}{r},
\end{equation}
\begin{eqnarray}\label{32}
H_{1PN} &=& \frac{1}{8}(3\eta-1)\textbf{p}^4-\frac{1}{2r}[(3+\eta)\textbf{p}^2 \nonumber \\
&& +\eta(\textbf{n}{\cdot}\textbf{p})^2]+\frac{1}{2r^2},
\end{eqnarray}
\begin{eqnarray}\label{33}
 H_{2PN} &=& \frac{1}{16}(1-5\eta+5\eta{^2})\textbf{p}^6 +\frac{1}{8r}[(5-20\eta \nonumber \\
&& -3\eta^2)\textbf{p}^4 -2\eta^2(\textbf{n}\cdot\textbf{p})^2\textbf{p}^2 \nonumber \\
&& -3\eta^2(\textbf{n}\cdot\textbf{p})^4] +\frac{1}{2r^2}[(5+8\eta)\textbf{p}^2  \nonumber \\
&& +3\eta(\textbf{n}\cdot\textbf{p})^2]-\frac{1}{4r^3}(1+3\eta).
\end{eqnarray}
$H_{SO}$ is a spin-orbit coupling contribution at 1.5 PN order
(Buonanno et al. 2006)
\begin{equation}\label{34}
H_{SO}=\frac{\eta}{r^{3}} \mathbf{L} \cdot \mathbf{S}_{\mathrm{eff}},
\end{equation}
where
\begin{equation}\label{37}
\mathbf{S}_{\mathrm{eff}}=\left(2+\frac{3}{2}{}\frac{1}{\gamma}\right)
\mathbf{S}_{1}+\left(2+\frac{3}{2} \gamma\right) \mathbf{S}_{2}.
\end{equation}
$H_{SS}$ is a spin--spin coupling contribution at 2 PN order
(Buonanno et al. 2006)
\begin{equation}\label{35}
H_{SS}=\frac{\eta}{2 r^{3}}\left[3\left(\mathbf{S}_{0} \cdot
\mathbf{n}\right)^{2}-\mathbf{S}_{0}^{2}\right],
\end{equation}
where
\begin{equation}\label{38}
\mathbf{S}_{0}=(1+\frac{1}{\gamma}) \mathbf{S}_{1}+(1+\gamma)
\mathbf{S}_{2}.
\end{equation}
Note that the Newton Wigner-Pryce spin supplementary condition
$\kappa = 0$ is considered (Mik\'{o}czi 2017).

The evolution of $(\bm{r},\bm{p})$ satisfies the canonical
equations
\begin{eqnarray}\label{39}
\frac{d \mathbf{r}}{d t} &=& \frac{\partial H}{\partial \mathbf{p}}, \\
\frac{d \mathbf{p}}{d t} &=& -\frac{\partial H}{\partial
\mathbf{r}}.\label{39b}
\end{eqnarray}
However, the two spins vary with time, in according with non-canonical
equations
\begin{equation}\label{40}
\frac{{\rm {d}} \mathbf{S}_{i}}{{\rm {d}} t}=\frac{\partial
H}{\partial \mathbf{S}_{i}} \times \mathbf{S}_{i}.
\end{equation}
Equations \eqref{39}-\eqref{40} determine four integrals of motion
in the system in \eqref{29}. The integrals are the Hamiltonian
\eqref{29} as an energy integral
\begin{equation}\label{29b}
E=H,
\end{equation}
and the total angular momentum vector
\begin{equation}\label{40b}
\mathbf{J}=\mathbf{L}+\mathbf{S}_{1}+ \mathbf{S}_{2}.
\end{equation}

Noticing the non-canonical Equation \eqref{40}, Wu $\&$ Xie (2010)
introduced a pair of canonical variables $(\theta_i, \xi_i)$ to
express each of the spins in the form
\begin{equation}\label{41}
\mathbf{S}_{i}=\left(\begin{array}{c}
\rho_{i} \cos \theta_{i} \\
\rho_{i} \sin \theta_{i} \\
\xi_{i}
\end{array}\right),
\end{equation}
where $\rho_{i} = \sqrt{\mathbf{S}_{i}^{2}-\xi_{i}^{2}}$.
Clearly, $\bm{S}_{i}$ is a two-dimensional vector with three
components. In this way, a ten-dimensional phase-space canonical
Hamiltonian with five degrees of freedom is obtained via
\begin{eqnarray}\label{42}
&& \mathcal{H}(x,y,z,\theta_{1},\theta_{2},p_{x},p_{y},p_{z},\xi_{1},\xi_{2}) \nonumber \\
&& = H_{O}(x,y,z,p_{x},p_{y},p_{z}) \nonumber \\
&& + H_{SO}(x,y,z,\theta_{1},\theta_{2},p_{x},p_{y},p_{z},\xi_{1},\xi_{2})\nonumber \\
&& + H_{SS}(x,y,z,\theta_{1},\theta_{2},\xi_{1},\xi_{2}).
\end{eqnarray}
Here, $\theta_i$ are generalized coordinates, and $\xi_i$ are
conjugate momenta. These satisfy the canonical equations
\begin{eqnarray}\label{43}
\frac{d \theta_i}{d t} &=& \frac{\partial \mathcal{H}}{\partial \xi_i}, \\
\frac{d \xi_i}{d t} &=& -\frac{\partial \mathcal{H}}{\partial
\theta_i}.\label{43b}
\end{eqnarray}

The EC method is also available for Equations \eqref{39},
\eqref{39b}, \eqref{43} and \eqref{43b}.

\subsection{Numerical tests}

In addition to the EC method, the aforementioned other algorithms  RK, IS,
and ES are also independently used. Letting the time step $h=0.1$,
and the mass ratio $\gamma=1$, we take the initial conditions
$x=40$, $y=z=p_{x}=p_{z}=0$, $p_{y}=\sqrt{(1-e)/x}$,
$\theta_{1}=\theta_{2}=\pi/4$, and $\xi_{1}=\xi_{2}=0.1$, where the
initial eccentricities are $e=0.0985$ for Orbit 1, and $e=0.7098$
for Orbit 2. In Figures 7 (a) and (b), the EC method almost gives
the machine double-precision to the energies of Orbits 1 and 2 in
an integration time of $t=1000$. When the integration spans this
time, and tends to $t=10^{7}$ corresponding to $10^{8}$ steps, a
slight secular drift in the energy errors occurs due to the
roundoff errors. Here, EC demonstrates the best level of accuracy, as compared
with the three other integrators, RK, IS, and ES. It is almost the same as the
high-precision method RKF89 in terms of the magnitude of energy errors and
the slope of error growth. The Hamiltonian errors for IS and ES
are larger than those for EC or RKF89, but have no secular
changes. Moreover, IS and ES can cause the errors of angular momentum to be
bounded for Orbit 1, with a smaller eccentricity in Figure 7 (c),
but to linearly grow for Orbit 2, with a larger eccentricity in
Figure 7 (d). The errors of angular momentum for EC are similar to
those for RK, and grow with time. This result is reasonable, because
EC conserves the energy rather than the angular momentum from the
theoretical viewpoint.

It can also be seen from Figures 7 (a) and (b) that the initial
eccentricities do not exert an explicit influence on the
Hamiltonian errors for the EC method, unlike the three other
schemes, RK, IS, and ES. To clearly show this, Figure 8 plots the
dependence of the Hamiltonian errors given by the  algorithms on
the initial eccentricities, where several  values of $\gamma$ are
given. The three methods RK, IS, and ES show no dramatic
differences in the Hamiltonian errors for smaller initial
eccentricities. As the initial eccentricity increases, the
Hamiltonian errors increase for the RK method. The IS and ES
schemes result in large errors for some large initial
eccentricities. However, the energy errors made by the EC method
are like those made by RKF89, and are not explicitly dependent on the
initial eccentricities. The numerical performance of these
algorithms does not depend on the mass ratio, $\gamma$.

When we take a larger step size, such as $h=1$, in Figure 9,
EC, like RKF89, exhibits good accuracy up to an integration of
$10^{9}$ steps, although this leads to a secular drift in the energy
errors. The errors of IS and ES remain bounded, but are several
orders of magnitude larger than those of EC. The accuracy of
EC for the larger step size in Figure 9 is similar to that for the
smaller step size in Figure 7, i.e., it is independent of the
selected step size. However, the accuracy of IS, ES, and RK is
closely related to the selected step size. Figure 10 provides
further information regarding the dependence of algorithmic
accuracy on the initial eccentricities for different choices of
step size, h. It is once again clear that the energy accuracies of
EC and RKF89 are not sensitive to or dependent on the choice
of step sizes, unlike those of IS, ES, and RK.

Although EC is more time-consuming than IS, ES, and RK
in Figure 11, its energy accuracy is better, regardless of
whether the step size is small or large. In particular, EC is
suitable for larger step sizes. Such larger step sizes not only
cause EC to behave with better accuracy, since the roundoff
errors are decreased, but also leads to a reduction in EC's
computational cost. As well as the use of appropriately larger
step sizes, no other methods are considered to control roundoff
errors in the EC method. In fact, several authors have recently
been concerned with the issue of roundoff errors, e.g., Rein $\&$
Spiegel (2015), and Wisdom (2018).

Briefly, the main conclusions to be drawn from Figures 7 -- 11 are
that the energy conservation in the EC method is independent of
mass ratio, time step, or initial eccentricity. The EC method deals
with many iterative computations, and is therefore more expensive
in terms of computational cost than the implicit scheme, IS.
Fortunately, the application of appropriately larger time steps to the
EC method does not affect computational accuracy, and is very
helpful in reducing computational cost.

\subsection{Chaotic dynamics}

In addition to the mass ratio, time step, and initial
eccentricity, the dynamical feature of orbits exerts no influence
on the performance of the EC method. As shown above, the EC
method exhibits virtually identical errors for Orbits 1 and 2 in
Figure 7. The regularity of Orbit 1, and the chaoticity of Orbit
2 are shown via the power spectra and FLIs in Figure 12.

We use the EC method to discuss the relation between the
chaoticity of orbits and the initial separation, $r=x$. Where the
mass ratio $\gamma=1$, and the initial eccentricity $e$ is
also given several values, the initial separation, $x$, runs from 20
to 60 in an interval of 1. The FLI is obtained for a given
initial separation after the integration time $t=3.5 \times
10^{4}$. FLI=5 is still represents the threshold between the ordered and
chaotic cases. The dependence of FLIs on the initial separations,
$x$, for different initial eccentricities, $e$, is plotted in Figure
13. An important result is that chaos occurs easily for smaller
initial separations with higher initial eccentricities. These
results can be observed quite clearly from the $x$--$e$ plane,
colored in different values of FLIs in Figure 14, and are
consistent with those of Hartl  $\&$ Buonanno (2005). In fact, the
occurrence of chaos is completely due to the spin--spin coupling
contribution in Equation (28). If the spin--spin coupling is
dropped, the system (21) is integrable and non-chaotic. When it is
included, the system (21) is non-integrable, and then chaos becomes
possible. For the case of smaller initial separations and higher
initial eccentricities, the spin--spin coupling effects become
larger. This is more likely to give rise to the occurrence of chaos.

In short, the EC method can provide reliable numerical
results with respect to the long-term evolution of spinning
compact binaries, comparable with those of the high-precision
algorithm RKF89. The FLI technique is a convenient tool for
finding chaos by scanning a two-dimensional space with
specific parameters and initial conditions.

\section{Summary}

By performing a suitable discretization-averaging of the
Hamiltonian canonical equations of a ten-dimensional phase-space
Hamiltonian system with five degrees of freedom, in this paper, we
have proposed an implicit nonsymplectic exact energy-preserving
integrator. The discretization-averaging involves each component
of the Hamiltonian gradient being approximately replaced with the
average of ten ratios of Hamiltonian difference terms to the
position or momentum increments. This approach confers a
second-order accuracy on the numerical solutions.

When the new energy-conserving method is applied to a
one-dimensional disordered discrete nonlinear Schr\"{o}dinger
equation, it exhibits good  long-term numerical performance in the
preservation of energy. This performance is independent of the
regular and chaotic behavior of orbits. The new method can provide
reliable numerical results to a problem over a long-term numerical
integration, comparable to those of the high-precision algorithm
RKF89. With the
aid of this numerical integrator and fast Lyapunov indicators, the
influence of the parameters on chaos can be studied. We have shown
numerically that $\beta$, rather than $\varepsilon_{m}$, makes
a significant contribution to the occurrence of chaos. No chaos can
exist if $\beta$ is too small.

When the newly proposed energy-conserving integrator solves
the post-Newtonian Hamiltonian system of spinning compact
binaries, it still works well for long-term numerical integrations.
This good long-term performance is not affected by the mass ratio,
time step, initial eccentricity, or the regular/chaotic dynamical
properties of orbits. Unfortunately, the new method is implicit, and
therefore expensive in terms of computational cost. However, the
use of appropriately large time steps makes it less time-consuming
and reduces roundoff errors. The new method, combined with the
technique of fast Lyapunov indicators, is highly effective in
identifying the dynamical transition from order to chaos when a
certain parameter or initial condition is varied. The results,
concluded based on a scan of the fast Lyapunov indicators in a
two-dimensional space, based on initial separation and eccentricity,
is as follows: a combination of small initial separations and high
initial eccentricities plays an important role in inducing chaos.

When roundoff errors are neglected, the new method has a
good long-term energy conservation, irrespective of time steps,
eccentricities of orbits, or the regularity/chaoticity of orbits. It
can provide reliable numerical results. Thus, it is worth
recommending this approach in order to simulate Hamiltonian
problems with a ten-dimensional phase space, if appropriately
large time steps are chosen.

\section*{Acknowledgments}

The authors are very grateful to the referee for valuable comments
and useful suggestions. This research has been supported by the
National Natural Science Foundation of China [grant Nos. 11973020
(C0035736), 11533004, 11663005, 11533003, and 11851304], the
Special Funding for Guangxi Distinguished Professors
(2017AD22006), and the National Natural Science Foundation of
Guangxi (Nos. 2018GXNSFGA281007, and 2019JJD110006).

\appendix
\section*{Appendix}

\section{Proof of numerical solutions accurate to the order of $h^2$}

Expanding each increment of the Hamiltonian between $q_{1(n+1)}$
and $q_{1(n)}$  in Equation \eqref{6} to second-order partial
derivatives at point $(0,0,0,0,0,0,0,0,0,0)$ in according with the Taylor
expansion, we rewrite Equation \eqref{6} as follows:
\begin{eqnarray}\label{a1}
q_{1(n+1)}-q_{1(n)} &=&
\frac{h/10}{p_{1(n+1)}-p_{1(n)}}[(H(00000 10000)-H(00000 00000))
+(H(10000 10000)  \nonumber \\
&&  -H(10000 00000)) +(H(00001 10001)-H(00001 00001)) +(H(11000 11000) \nonumber \\
&& -H(11000 01000)) +(H(00011 10011)-H(00011 00011))
+(H(11100 11100) \nonumber \\
&& -H(11100 01100)) +(H(00111 10111) -H(00111 00111))
+(H(11110 11110) \nonumber \\
&& -H(11110 01110)) +(H(01111 11111)-H(01111 01111))
+(H(11111 11111) \nonumber \\
&& -H(11111 01111))] \nonumber \\
 &=&
\frac{h}{10} [(\frac{\partial H}{\partial
p_1}+\frac{1}{2}\frac{\partial^2 H}{\partial p^2_1}\Delta
p_1)+(\frac{\partial }{\partial
p_1}H(10000 00000)+\frac{1}{2}\frac{\partial^2 H}{\partial
p^2_1}\Delta
p_1) \nonumber \\
&& +(\frac{\partial }{\partial
p_1}H(00001 00001)+\frac{1}{2}\frac{\partial^2 H}{\partial
p^2_1}\Delta p_1) +(\frac{\partial }{\partial p_1}H(11000 01000)
+\frac{1}{2}\frac{\partial^2 H}{\partial
p^2_1}\Delta p_1)\nonumber \\
&& +(\frac{\partial }{\partial
p_1}H(00011 00011)+\frac{1}{2}\frac{\partial^2 H}{\partial
p^2_1}\Delta p_1) +(\frac{\partial }{\partial
p_1}H(11100 01100)+\frac{1}{2}\frac{\partial^2 H}{\partial
p^2_1}\Delta p_1)\nonumber \\
&& +(\frac{\partial }{\partial
p_1}H(00111 00111)+\frac{1}{2}\frac{\partial^2 H}{\partial
p^2_1}\Delta p_1) +(\frac{\partial }{\partial
p_1}H(11110 01110)+\frac{1}{2}\frac{\partial^2 H}{\partial
p^2_1}\Delta p_1)\nonumber \\
&& +(\frac{\partial }{\partial
p_1}H(01111 01111)+\frac{1}{2}\frac{\partial^2 H}{\partial
p^2_1}\Delta p_1) +(\frac{\partial }{\partial
p_1}H(11111 01111)+\frac{1}{2}\frac{\partial^2 H}{\partial
p^2_1}\Delta p_1)]\nonumber \\
 &=&
h\frac{\partial H}{\partial p_1}+\frac{h}{2}\frac{\partial^2
H}{\partial p^2_1}\Delta p_1+\frac{h}{10}[\frac{\partial^2}{\partial p_1\partial q_1}\Delta q_1
+(\frac{\partial^2}{\partial p_1\partial q_5}\Delta q_5+\frac{\partial^2}{\partial p_1\partial p_5}\Delta p_5) \nonumber \\
&& +(\frac{\partial^2}{\partial p_1\partial q_1}\Delta q_1
+\frac{\partial^2}{\partial p_1\partial q_2}\Delta q_2
+\frac{\partial^2}{\partial p_1\partial p_2}\Delta p_2) \nonumber \\
&& +(\frac{\partial^2}{\partial p_1\partial q_4}\Delta q_4
+\frac{\partial^2}{\partial p_1\partial q_5}\Delta q_5
+\frac{\partial^2}{\partial p_1\partial p_4}\Delta p_4
+\frac{\partial^2}{\partial p_1\partial p_5}\Delta p_5) \nonumber \\
&& +(\frac{\partial^2}{\partial p_1\partial q_1}\Delta q_1
+\frac{\partial^2}{\partial p_1\partial q_2}\Delta q_2
+\frac{\partial^2}{\partial p_1\partial q_3}\Delta q_3
+\frac{\partial^2}{\partial p_1\partial p_2}\Delta p_2
+\frac{\partial^2}{\partial p_1\partial p_3}\Delta p_3) \nonumber \\
&& +(\frac{\partial^2}{\partial p_1\partial q_3}\Delta q_3
+\frac{\partial^2}{\partial p_1\partial q_4}\Delta q_4
+\frac{\partial^2}{\partial p_1\partial q_5}\Delta q_5
+\frac{\partial^2}{\partial p_1\partial p_3}\Delta p_3
+\frac{\partial^2}{\partial p_1\partial p_4}\Delta p_4
+\frac{\partial^2}{\partial p_1\partial p_5}\Delta p_5) \nonumber \\
&& +(\frac{\partial^2}{\partial p_1\partial q_1}\Delta q_1
+\frac{\partial^2}{\partial p_1\partial q_2}\Delta q_2
+\frac{\partial^2}{\partial p_1\partial q_3}\Delta q_3
+\frac{\partial^2}{\partial p_1\partial q_4}\Delta q_4
+\frac{\partial^2}{\partial p_1\partial p_2}\Delta p_2
+\frac{\partial^2}{\partial p_1\partial p_3}\Delta p_3 \nonumber \\
&& +\frac{\partial^2}{\partial p_1\partial p_4}\Delta p_4)
+(\frac{\partial^2}{\partial p_1\partial q_2}\Delta q_2
+\frac{\partial^2}{\partial p_1\partial q_3}\Delta q_3
+\frac{\partial^2}{\partial p_1\partial q_4}\Delta q_4
+\frac{\partial^2}{\partial p_1\partial q_5}\Delta q_5
+\frac{\partial^2}{\partial p_1\partial p_2}\Delta p_2 \nonumber \\
&& +\frac{\partial^2}{\partial p_1\partial p_3}\Delta p_3
+\frac{\partial^2}{\partial p_1\partial p_4}\Delta p_4
+\frac{\partial^2}{\partial p_1\partial p_5}\Delta p_5)
+(\frac{\partial^2}{\partial p_1\partial q_1}\Delta q_1
+\frac{\partial^2}{\partial p_1\partial q_2}\Delta q_2
+\frac{\partial^2}{\partial p_1\partial q_3}\Delta q_3 \nonumber \\
&& +\frac{\partial^2}{\partial p_1\partial q_4}\Delta q_4
+\frac{\partial^2}{\partial p_1\partial q_5}\Delta q_5
+\frac{\partial^2}{\partial p_1\partial p_2}\Delta p_2
+\frac{\partial^2}{\partial p_1\partial p_3}\Delta p_3
+\frac{\partial^2}{\partial p_1\partial p_4}\Delta p_4+\frac{\partial^2}{\partial p_1\partial p_5}\Delta p_5)]H
\nonumber \\
&=& h\frac{\partial H}{\partial p_1}+\frac{h}{2}\frac{\partial
}{\partial  p_1}\sum^{5}_{j=1}(\frac{\partial }{\partial
q_j}\Delta q_j+{\partial  p_j}\Delta p_j)H+\mathcal{O}(h^3),
\end{eqnarray}
where $H=H(0000000000)$, and the position and momentum increments
are $\Delta q_j=q_{j(n+1)}-q_{j(n)}$ and $\Delta
p_j=p_{j(n+1)}-p_{j(n)}$ on the right-hand side of the above
equation. In a similar way, the other position and momentum
increments, including the position increment of $q_1$, are
expressed as
\begin{eqnarray}\label{a2}
\Delta q_i &=& h\frac{\partial }{\partial
p_i}H(0000000000)+\frac{h}{2}\frac{\partial }{\partial
p_i}\sum^{5}_{j=1}(\frac{\partial }{\partial q_j}\Delta
q_j+{\partial  p_j}\Delta p_j)H(0000000000)+\mathcal{O}(h^3),
\\ \Delta p_i &=& -h\frac{\partial }{\partial
q_i}H(0000000000)-\frac{h}{2}\frac{\partial }{\partial
q_i}\sum^{5}_{j=1}(\frac{\partial }{\partial q_j}\Delta
q_j+{\partial  p_j}\Delta p_j)H(0000000000)+\mathcal{O}(h^3),
\end{eqnarray}
where $i=1,\cdots,5$. The position and momentum increments on the
right-hand sides of Equations (A2) and (A3) take the first terms
\begin{eqnarray}\label{a3}
\Delta q_j &\approx& h\frac{\partial }{\partial p_i}H(0000000000)
\sim
\mathcal{O}(h), \\
\Delta p_j &\approx& -h\frac{\partial }{\partial q_i}H(0000000000)
\sim \mathcal{O}(h).
\end{eqnarray}
Substituting Equations (A4) and (A5) into Equations (A2) and (A3),
we obtain the numerical solutions
\begin{eqnarray}\label{a4}
RK: & & q_{i(n+1)} = p_{i(n)} +h\frac{\partial }{\partial
p_i}H(0000000000)+\frac{h}{2}\frac{\partial }{\partial
p_i}\sum^{5}_{j=1}(\frac{\partial }{\partial q_j}\Delta
q_j+{\partial  p_j}\Delta p_j)H(0000000000)+\mathcal{O}(h^3),
\\ & & p_{i(n+1)} = p_{i(n)}-h\frac{\partial }{\partial
q_i}H(0000000000)-\frac{h}{2}\frac{\partial }{\partial
q_i}\sum^{5}_{j=1}(\frac{\partial }{\partial q_j}\Delta
q_j+{\partial  p_j}\Delta p_j) H(0000000000)+\mathcal{O}(h^3).
\end{eqnarray}
It is clear that the numerical solutions are accurate to the order
of $h^2$.

In fact, the numerical solutions (A6) and (A7) are those given by
the refined Euler method, i.e., the second-order Runge-Kutta (RK)
method. They are also obtained from the second-order implicit
trapezoidal formula
\begin{eqnarray}\label{a5}
TR: && q_{i(n+1)}= q_{i(n)} +\frac{h}{2}(\frac{\partial }{\partial
p_i}H(0000000000)+\frac{\partial }{\partial p_i}H(1111111111)), \\
&& p_{i(n+1)}= p_{i(n)} -\frac{h}{2}(\frac{\partial }{\partial
q_i}H(0000000000)+\frac{\partial }{\partial q_i}H(1111111111)),
\end{eqnarray}
or the second-order implicit midpoint rule (Feng 1986; Zhong et
al. 2010; Mei et al. 2013a)
\begin{eqnarray}\label{a5}
IS: && q_{i(n+1)}= q_{i(n)} +h\frac{\partial }{\partial p_i}
H(\frac{0+1}{2}\frac{0+1}{2}\frac{0+1}{2}\frac{0+1}{2}\frac{0+1}{2}\frac{0+1}{2}\frac{0+1}{2}\frac{0+1}{2}\frac{0+1}{2}\frac{0+1}{2}), \\
&& p_{i(n+1)}= p_{i(n)} -h\frac{\partial }{\partial
q_i}H(\frac{0+1}{2}\frac{0+1}{2}\frac{0+1}{2}\frac{0+1}{2}\frac{0+1}{2}\frac{0+1}{2}\frac{0+1}{2}\frac{0+1}{2}\frac{0+1}{2}\frac{0+1}{2}),
\end{eqnarray}
where the solutions $q_{i(n+1)}$ and $p_{i(n+1)}$ in the
Hamiltonians $H(1111111111)$ and $H(\frac{0+1}{2}\cdots)$ take the
first and second terms on the right-hand side of Equations (A6)
and (A7), and $H(1111111111)$ and $H(\frac{0+1}{2}\cdots)$ are
expanded to the order of $h$. Although the four algorithms,
including the new method in Equations (6)-(15) for RK, TR, and IS,
have the same order, they are different in terms of numerical performance.
RK does not conserve the energy integral. IS is symplectic, and
shows no secular drift in energy errors. TR is the same as IS when
it is used to solve a linear Hamiltonian system, but is not
symplectic for a nonlinear Hamiltonian system (Feng $\&$ Qin
2009). The new method in Equations (6)-(15) is exactly
energy-conserving, theoretically, and moreover, is not symplectic.


\begin{figure*}
\center{
\includegraphics[scale=0.3]{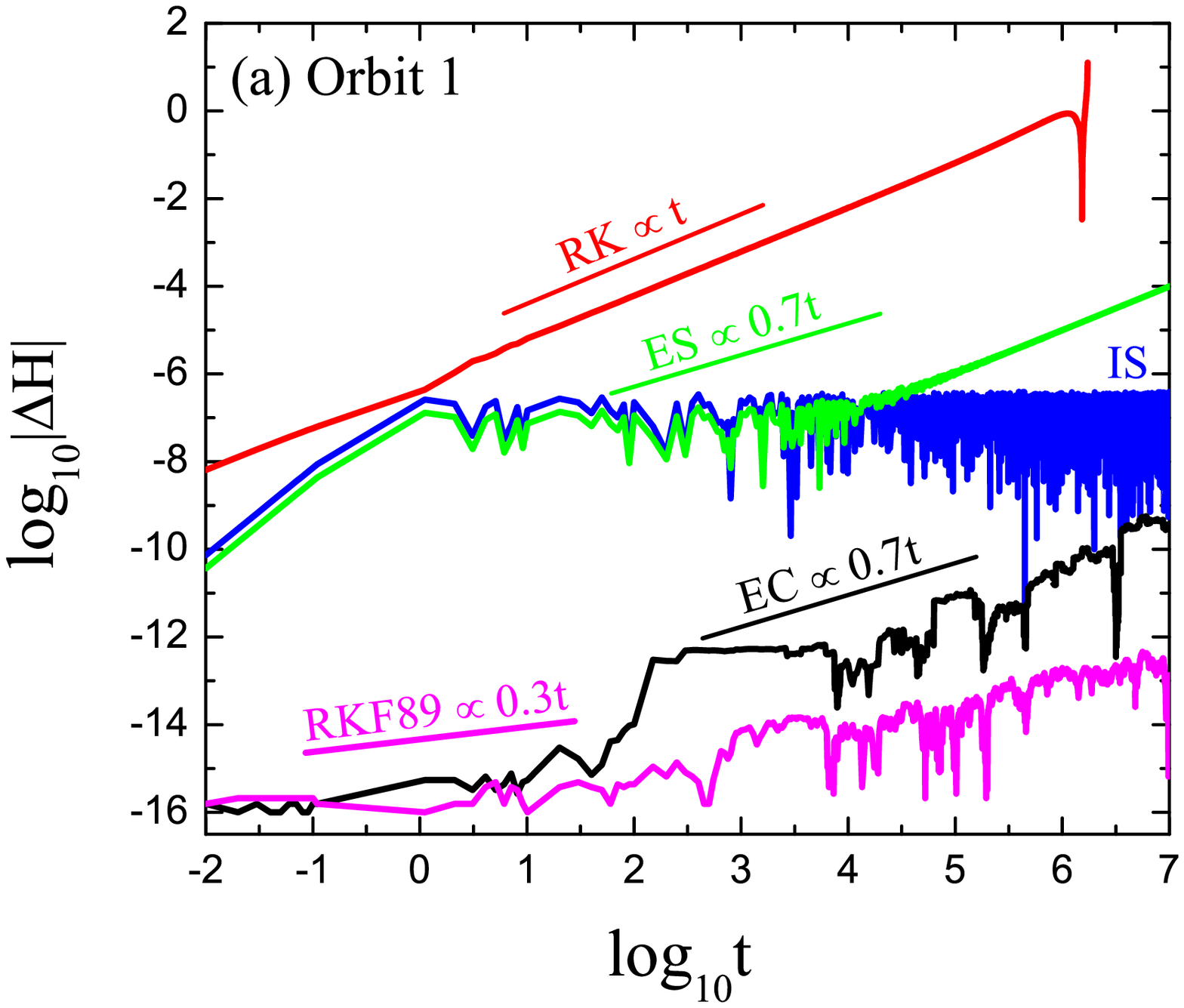}
\includegraphics[scale=0.3]{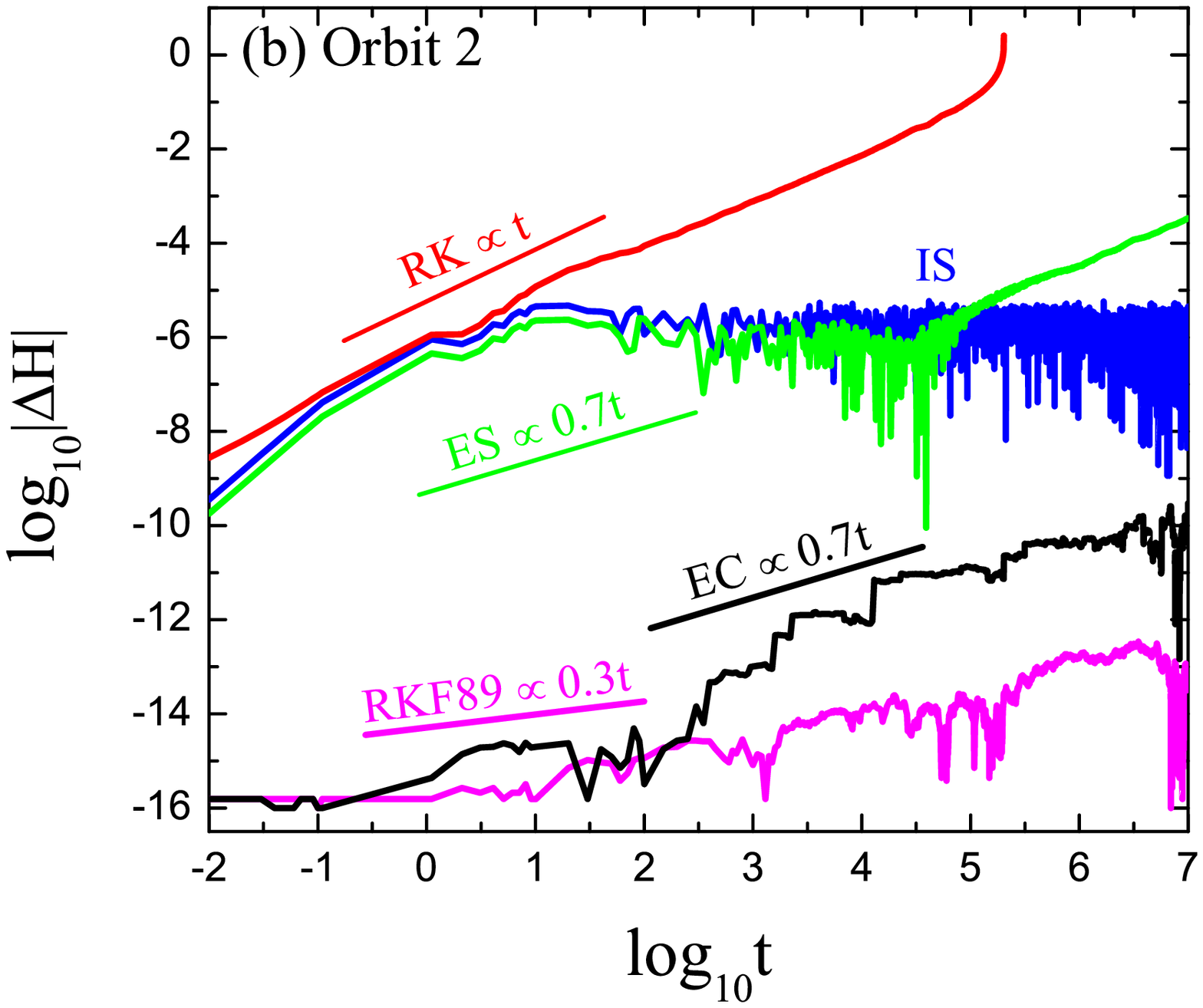}
\includegraphics[scale=0.3]{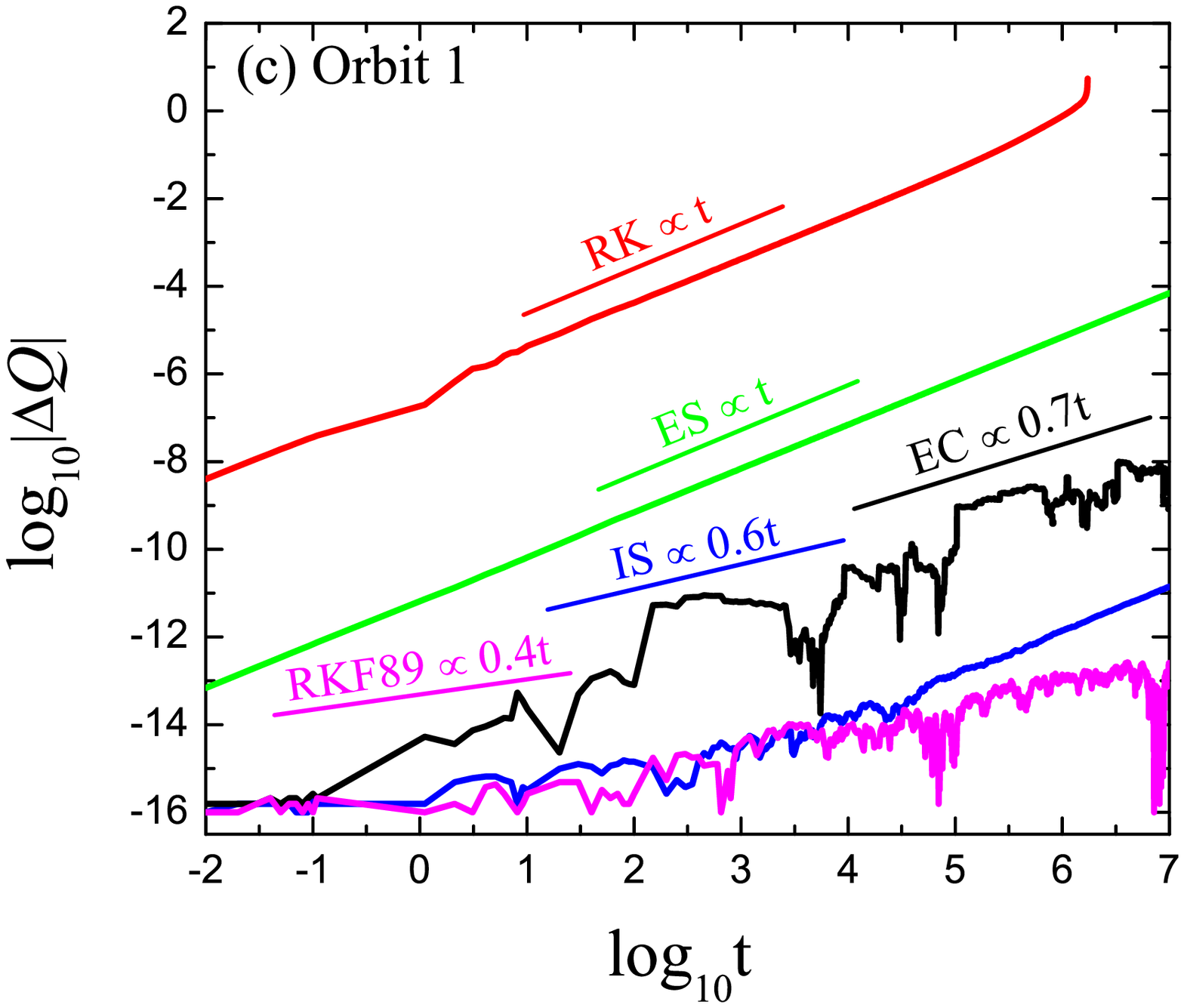}
\includegraphics[scale=0.3]{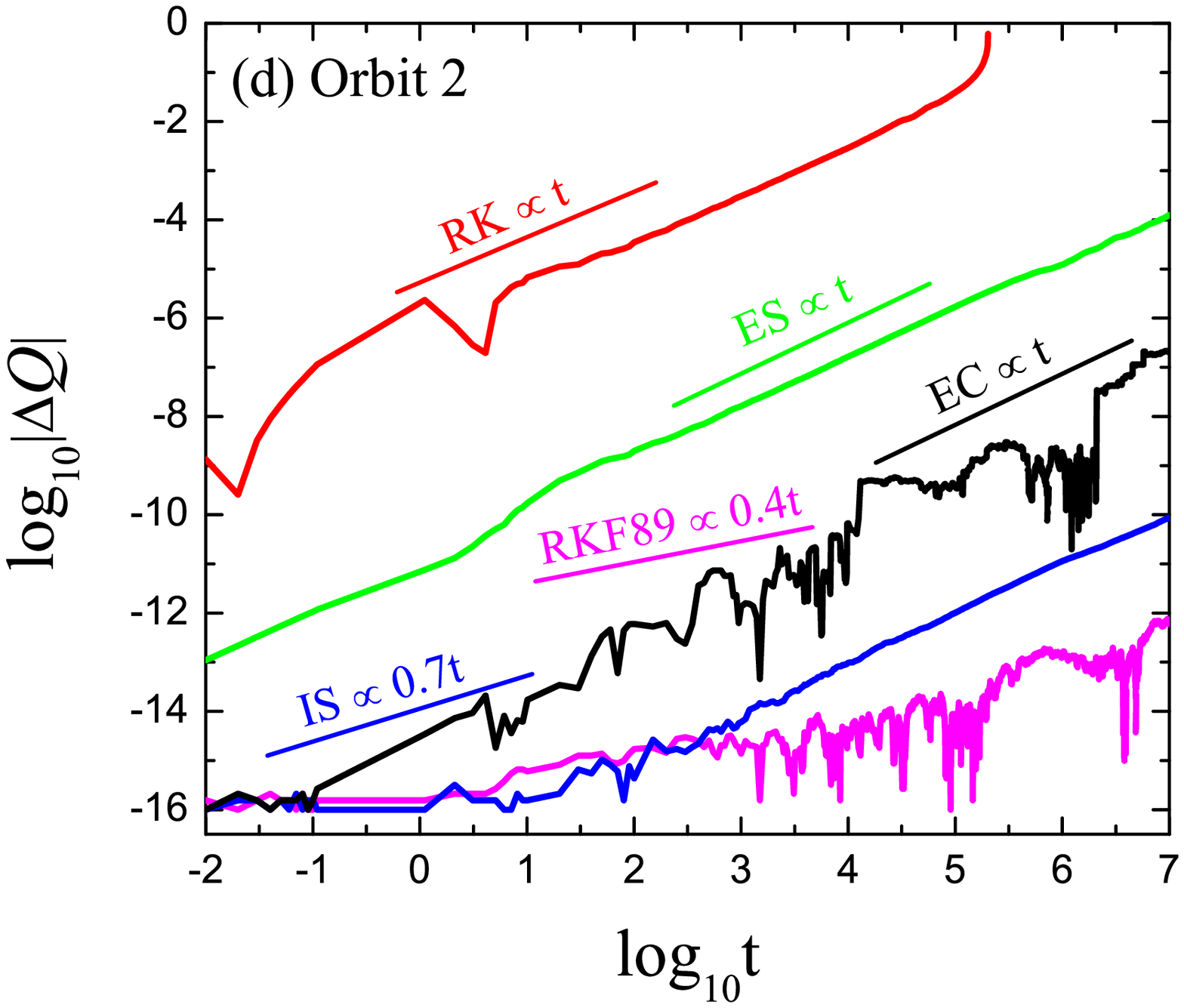}
\caption{Errors of the Hamiltonian and norm, $\Delta H$ and
$\Delta Q$, for the five algorithms solving Orbit 1 and Orbit 2 in
the ten-dimensional phase-space DDNLS system. The time step is
$h=0.01$.  Orbits 1 and 2 have the same initial conditions, where
$q_{i}=(6-i)/10$ and $p_{i}=0$. The two orbits take the same
values, $\varepsilon_{m}=\varepsilon=0$. $\beta = 1$ for Orbit 1,
and $\beta = 10$  for Orbit 2. The slope of the error growth with
time is marked for each algorithm. When the integration time
$t=10^7$ corresponds to $10^9$ integration steps, the implicit
symplectic midpoint method IS causes the Hamiltonian errors to
have no secular drift. Although the Hamiltonian errors of the new
energy-conserving method EC, yield a slightly secular growth, they
are smaller than those of IS, and are approximate to those of the
high-precision method, RKF89.}} \label{fig1}
\end{figure*}

\begin{figure*}
\center{
\includegraphics[scale=0.4]{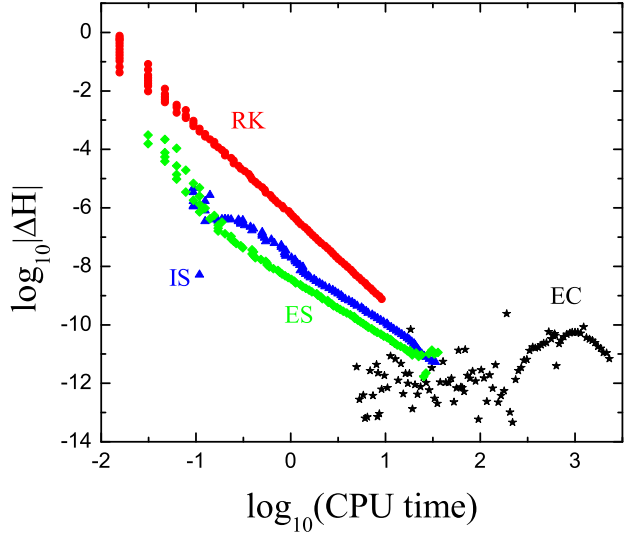}
\caption{Efficiencies of algorithms EC, IS, ES, and RK. In fact,
the efficiency describes the relation between the maximum
Hamiltonian error and CPU time (unit: second) for a given step
size. Each orbit has the same initial conditions,
where $q_{i}=(6-i)/10$ and $p_{i}=(6-i)/100$,
$\varepsilon_{m}=\varepsilon=0$, and $\beta = 1$. The maximum error
is obtained after the integration time $t=10^{4}$. The time steps
are fixed for each algorithm, but the points correspond to
different time steps $h=0.05/1.07226722^{k-1}$, where
$k=1,2,\cdots,100$. Shorter CPU times correspond to larger step
sizes, and longer CPU times correspond to smaller step sizes.
Although EC has the poorest efficiency, its accuracy does not
depend on the choice of step sizes.}} \label{fig2}
\end{figure*}

\begin{figure*}
\center{
\includegraphics[scale=0.2]{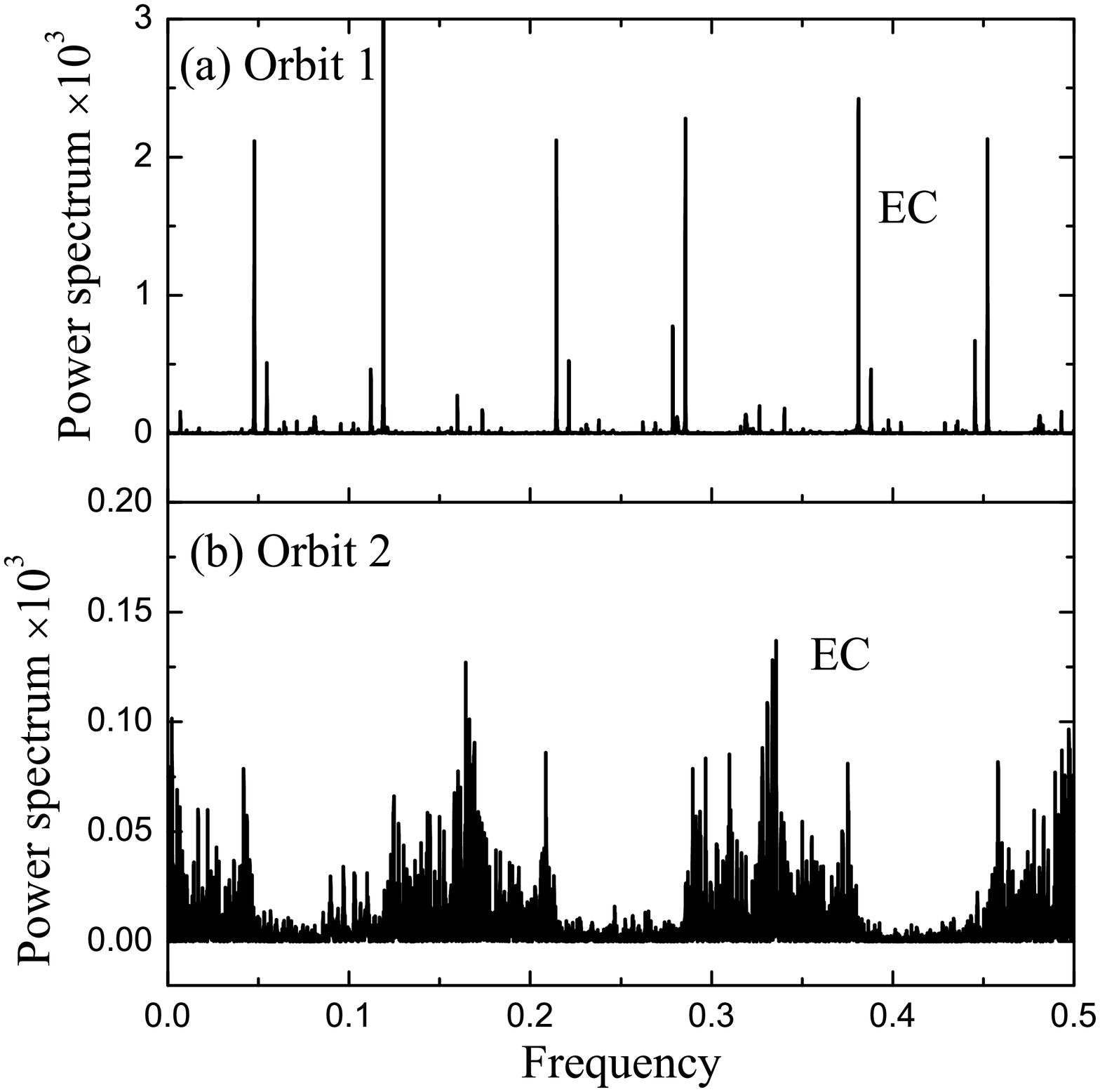}
\includegraphics[scale=0.2]{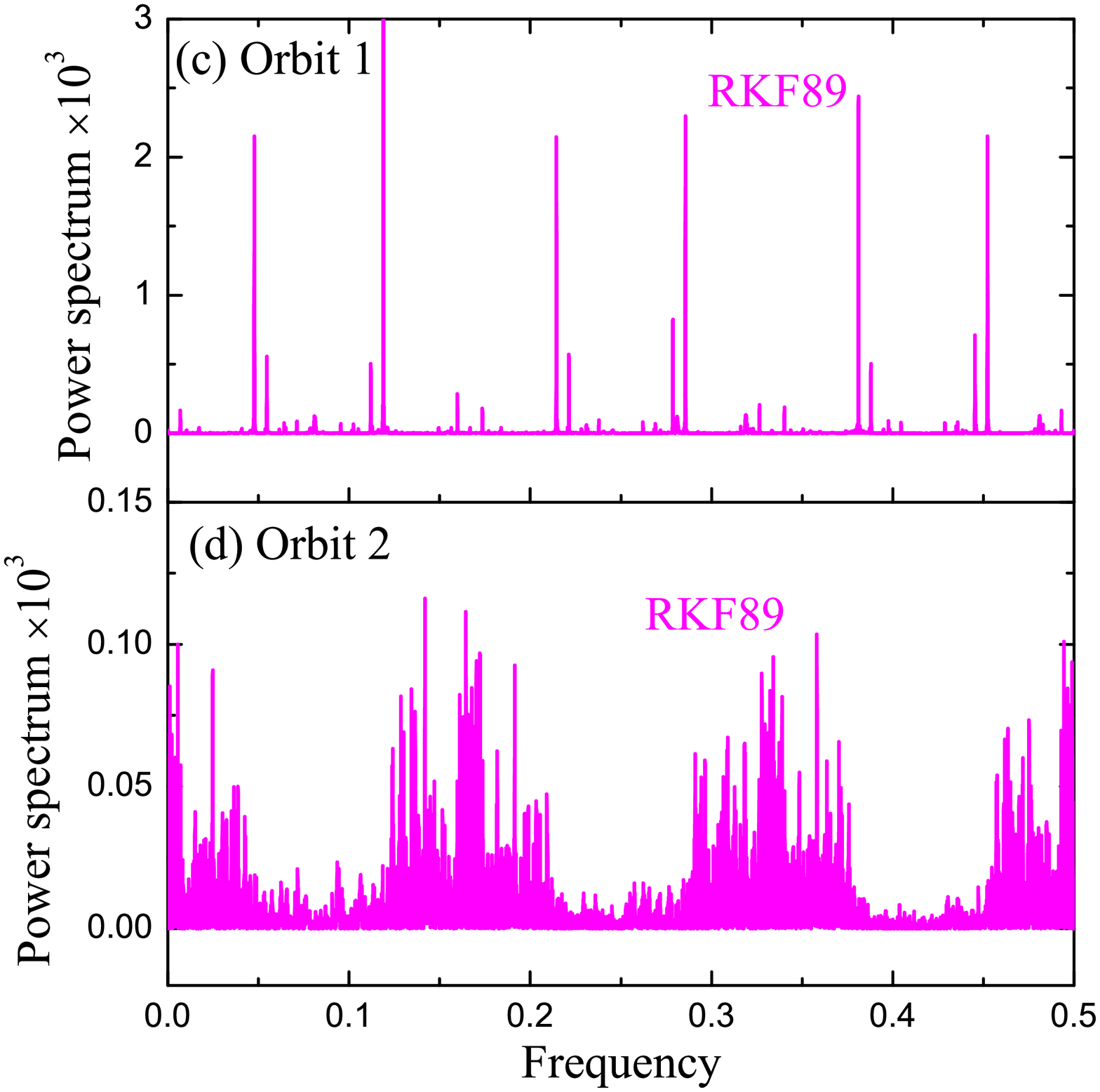}
\includegraphics[scale=0.2]{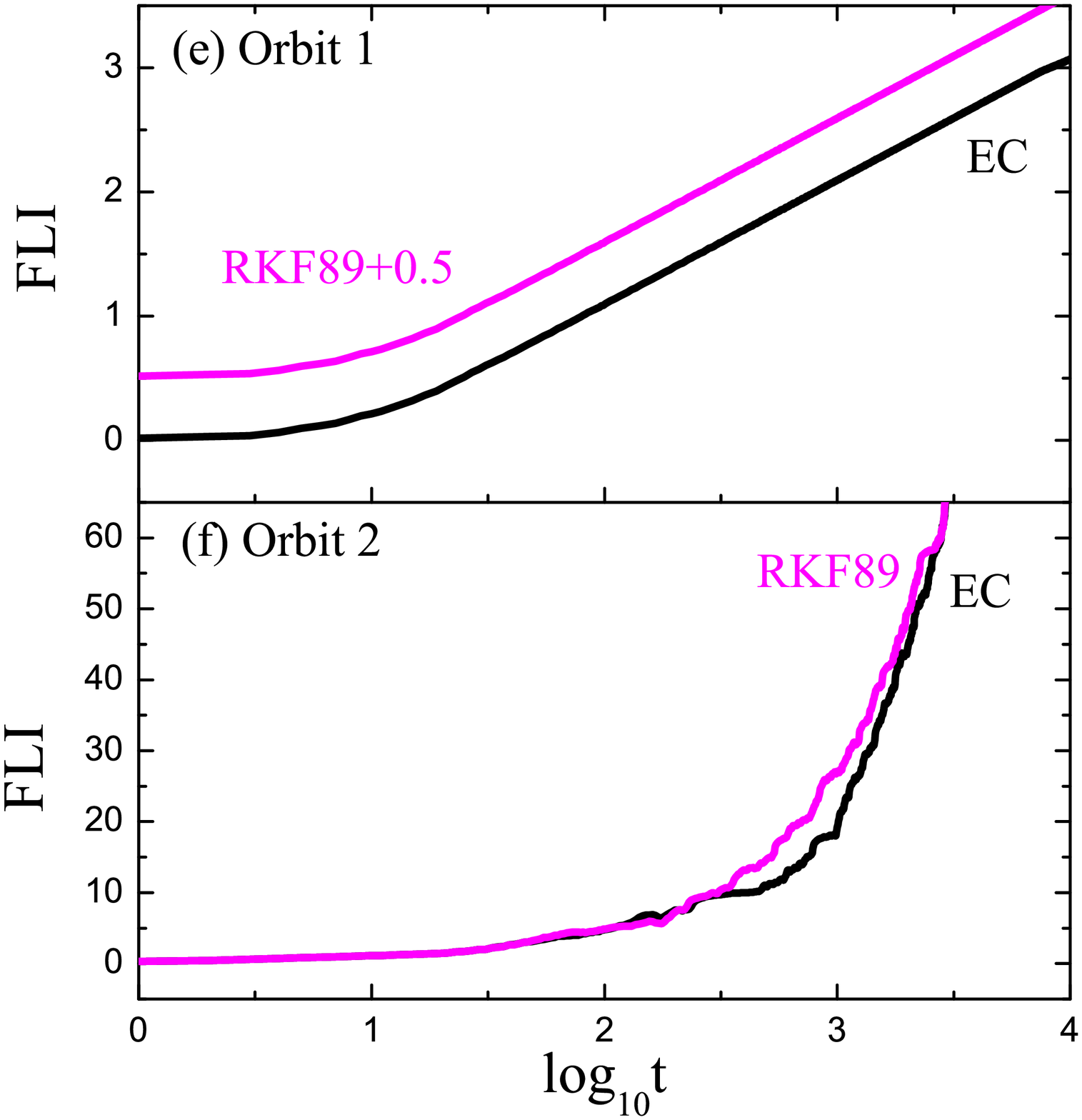}
\caption{Power spectra and fast Lyapunov indicators (FLIs)  of
Orbits 1 and 2, given by methods EC and RKF89. They show that
Orbit 1 is ordered, and Orbit 2 is chaotic. EC and RKF89 have the
same results. Note that EC and RKF89 approximately coincide in
panel (e).}} \label{fig3}
\end{figure*}

\begin{figure*}
\center{
\includegraphics[scale=0.3]{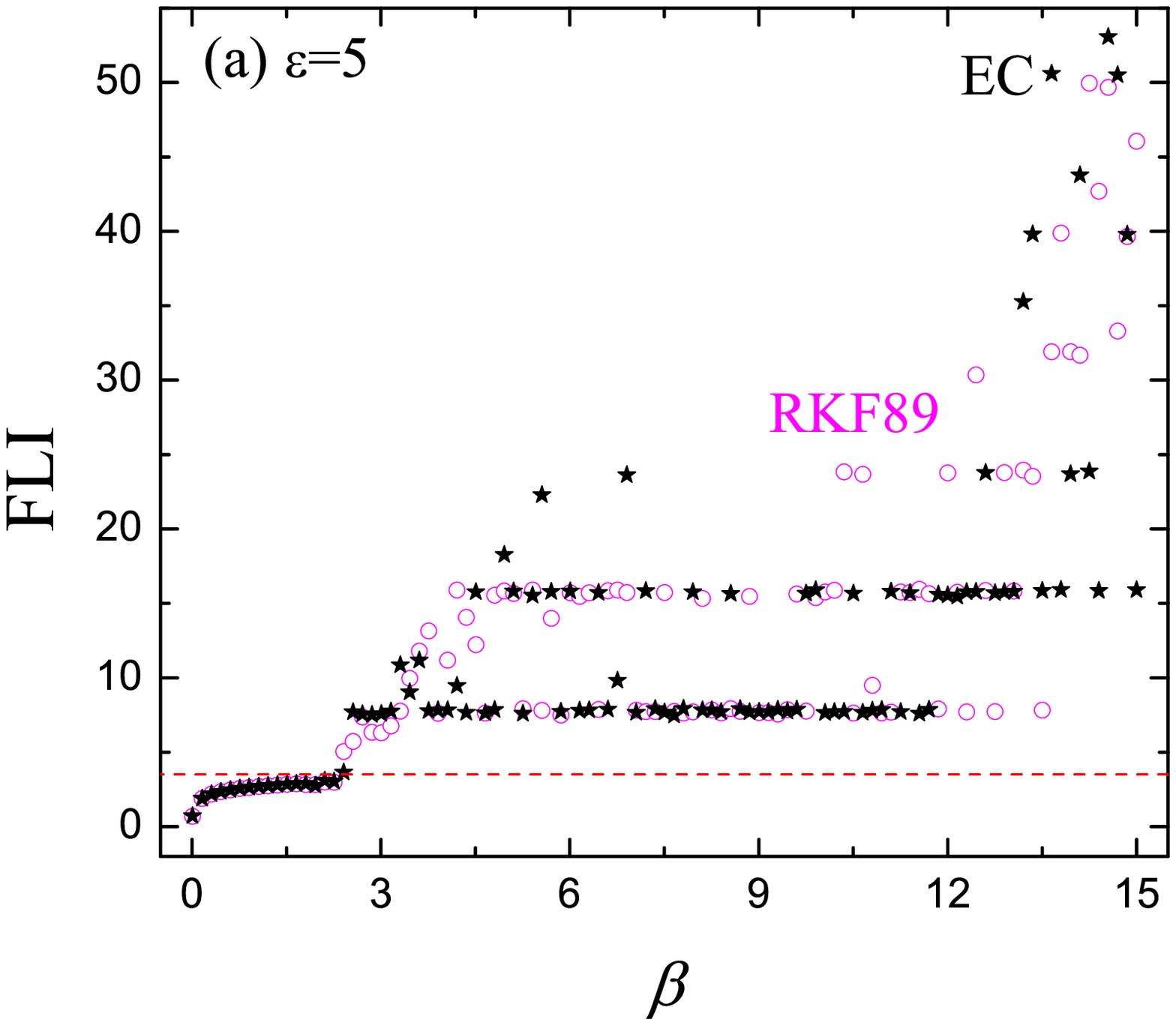}
\includegraphics[scale=0.3]{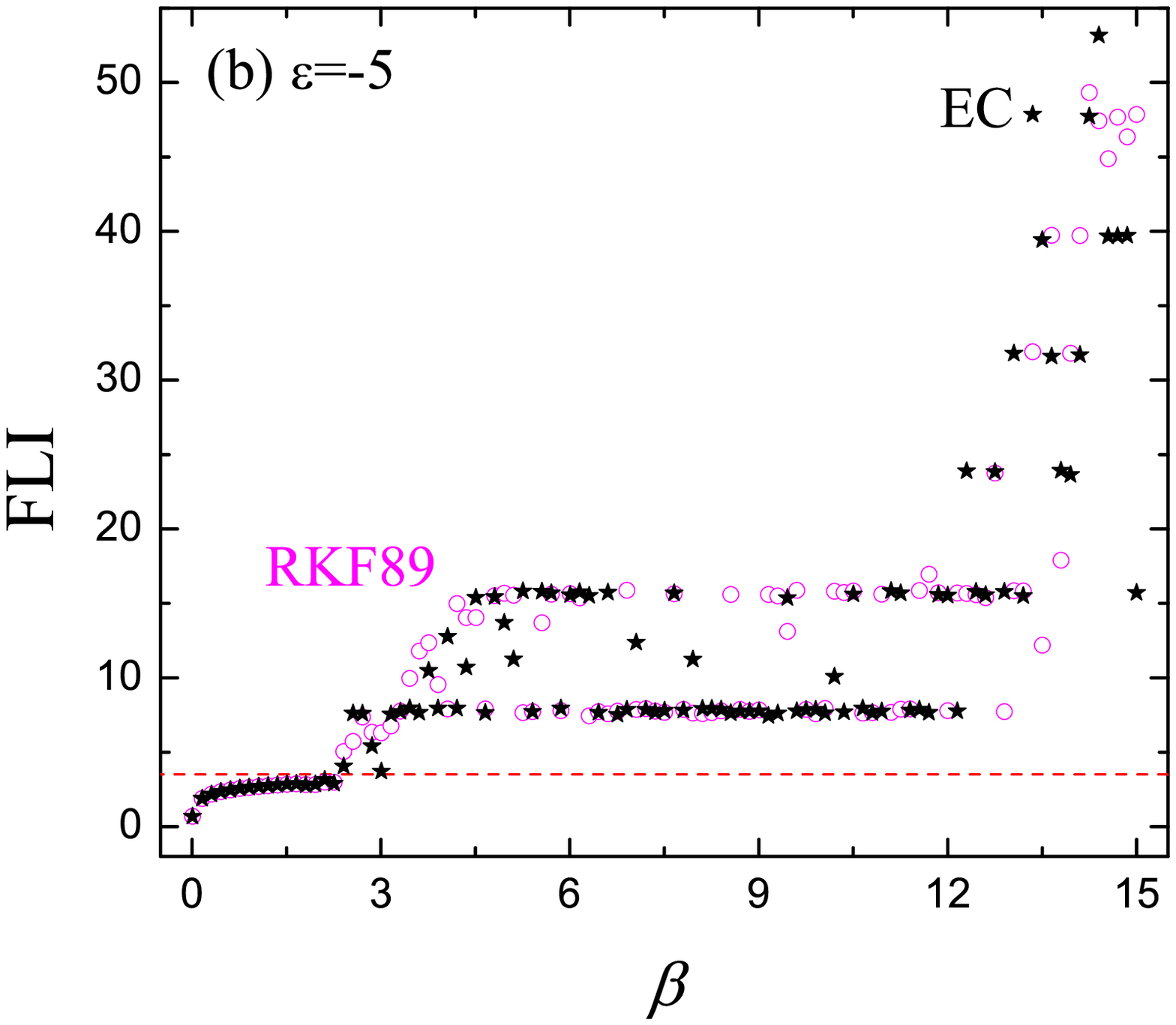}
\includegraphics[scale=0.3]{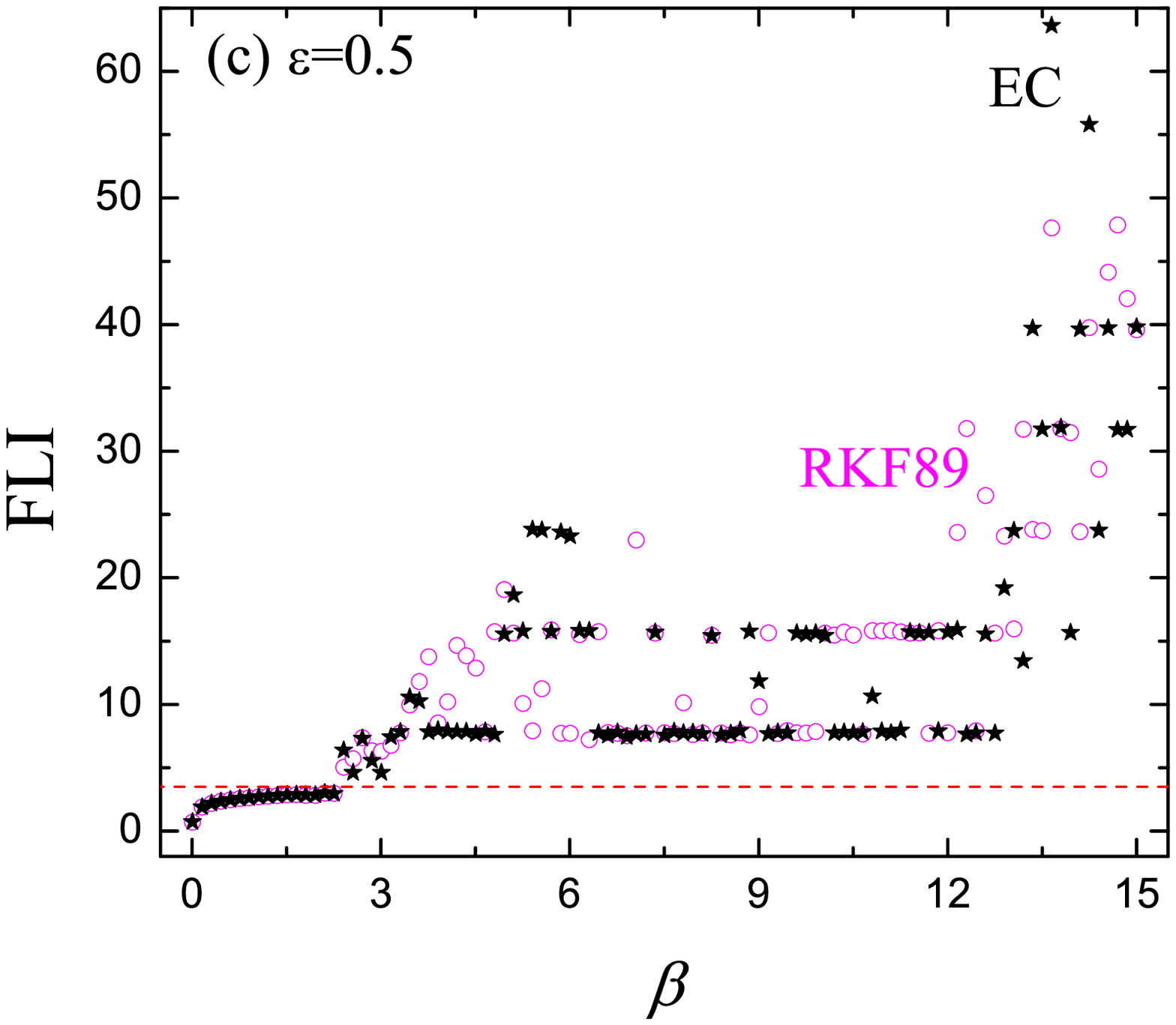}
\includegraphics[scale=0.3]{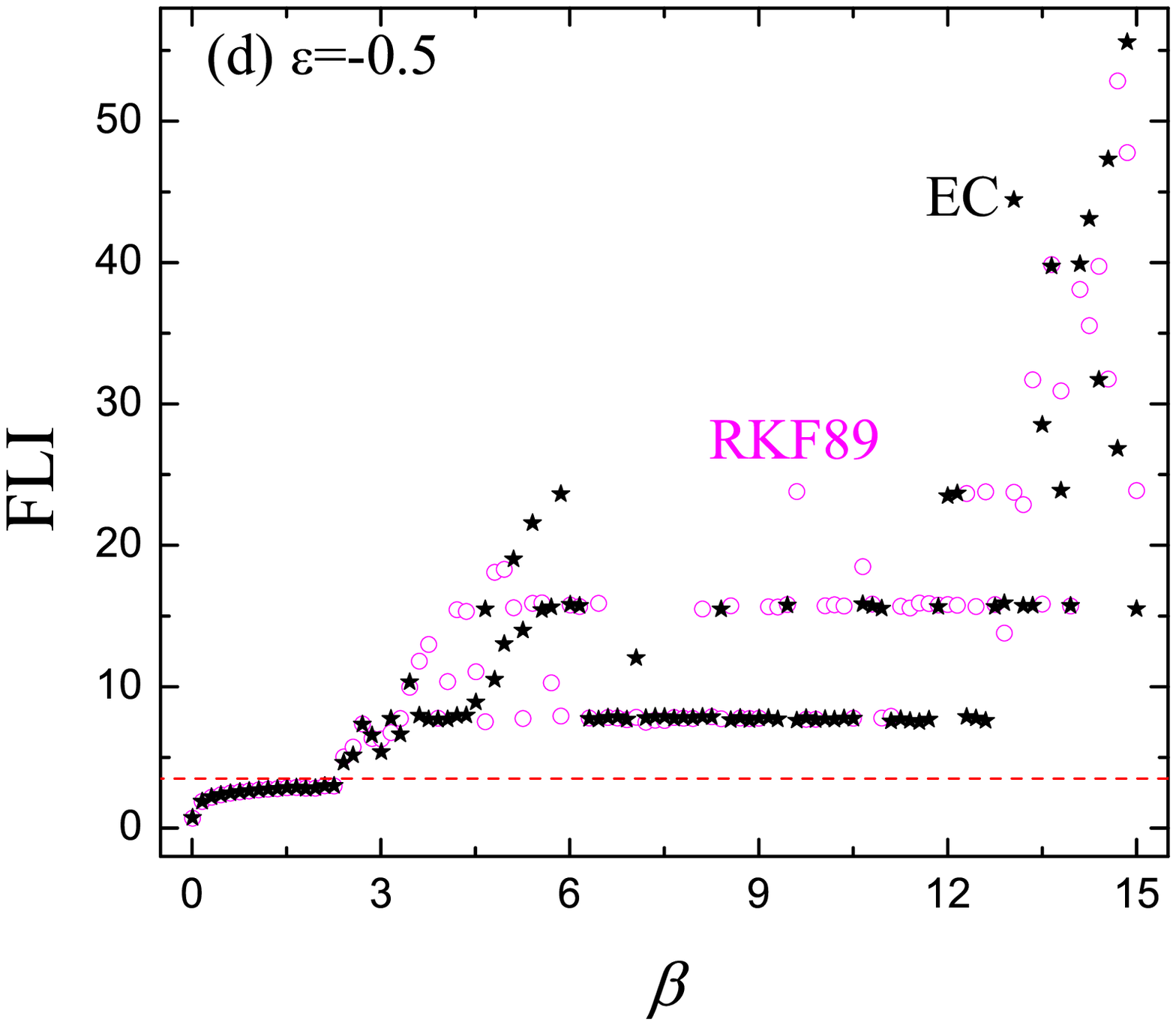}
\caption{Dependence of FLIs on parameter $\beta$. Each orbit has
the same initial variables, $q_{i}=(6-i)/10$ and $p_{i}=(6-i)/100$.
Given a value of parameter $\beta$, the FLI is obtained after the
integration time $t=3800$. A value of $5$ represents the threshold for FLIs, between
ordered and chaotic cases. FLI$>$5 indicates the chaoticity, and
FLI$\leq$5 indicates the regularity. Both EC and RKF89 give the same
results.}} \label{fig4}
\end{figure*}

\begin{figure*}
\center{
\includegraphics[scale=0.3]{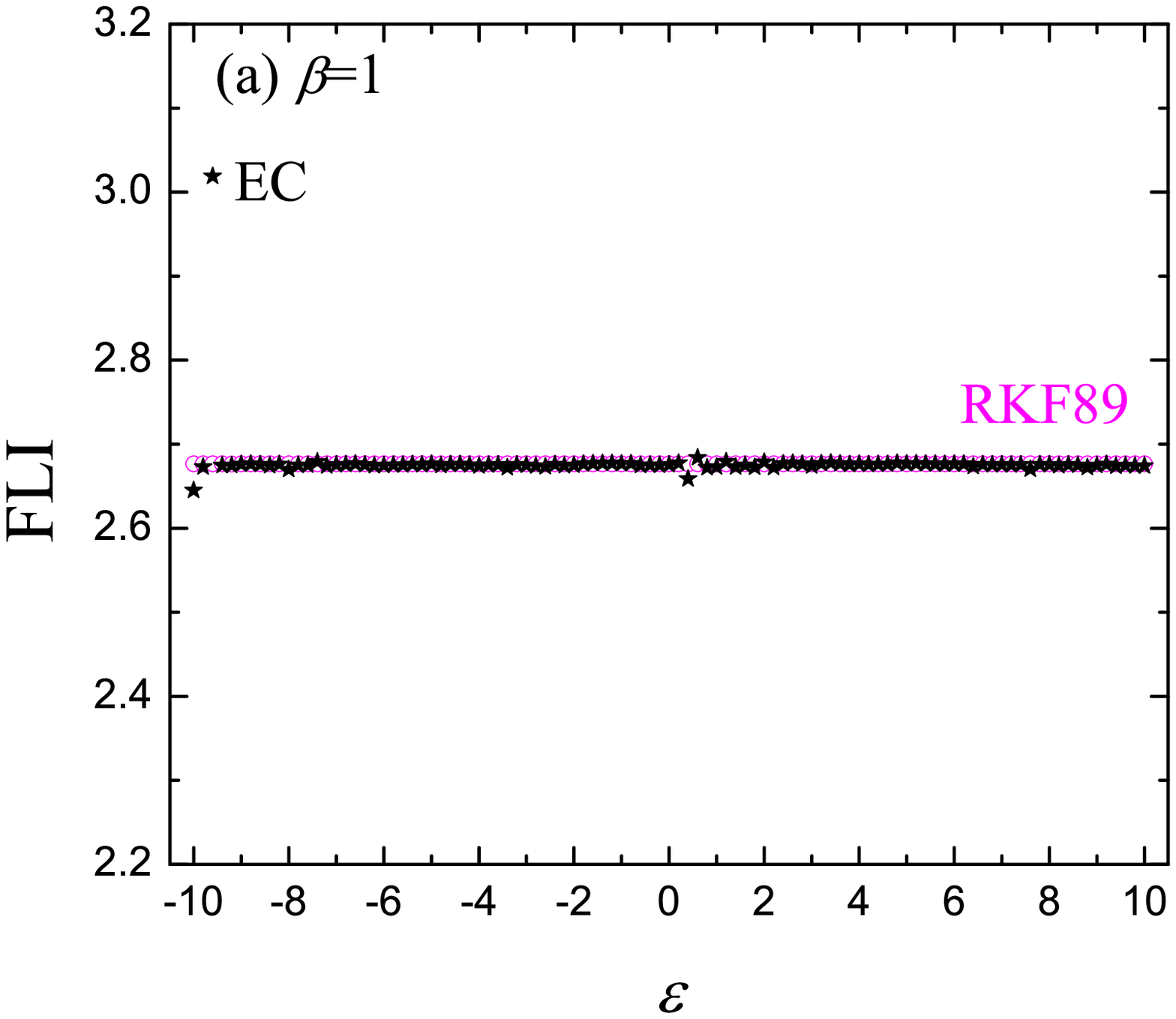}
\includegraphics[scale=0.3]{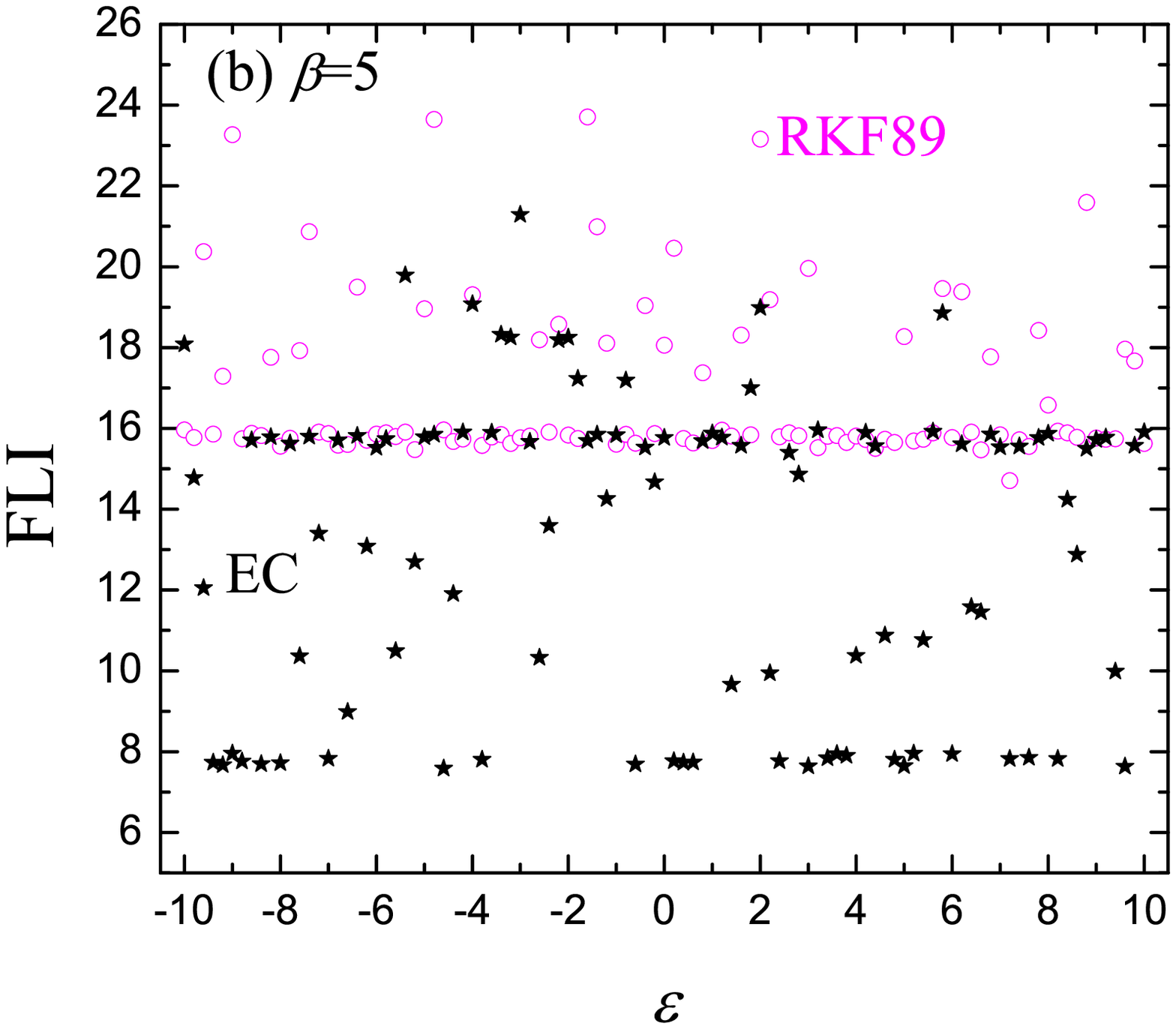}
\includegraphics[scale=0.3]{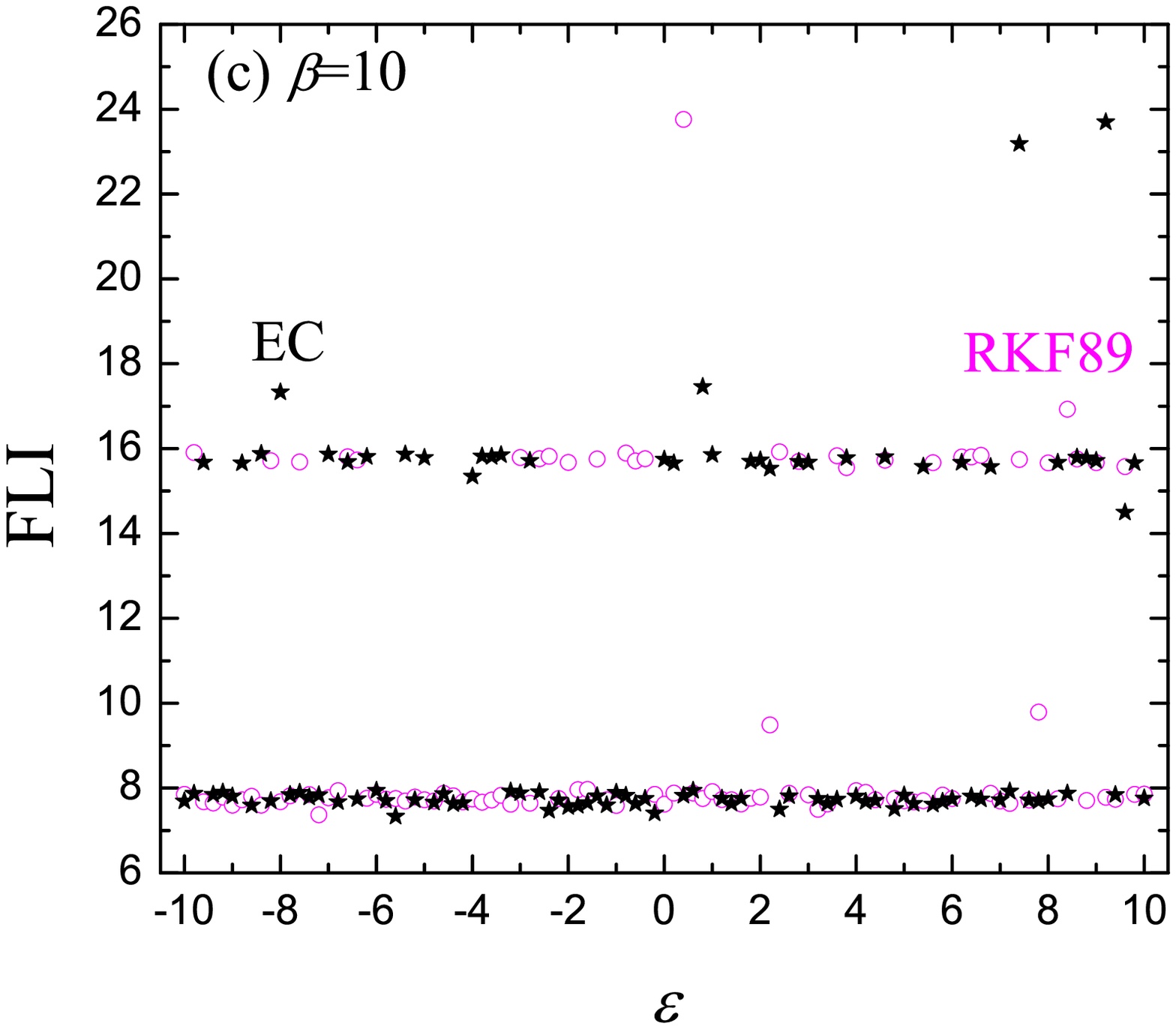}
\includegraphics[scale=0.3]{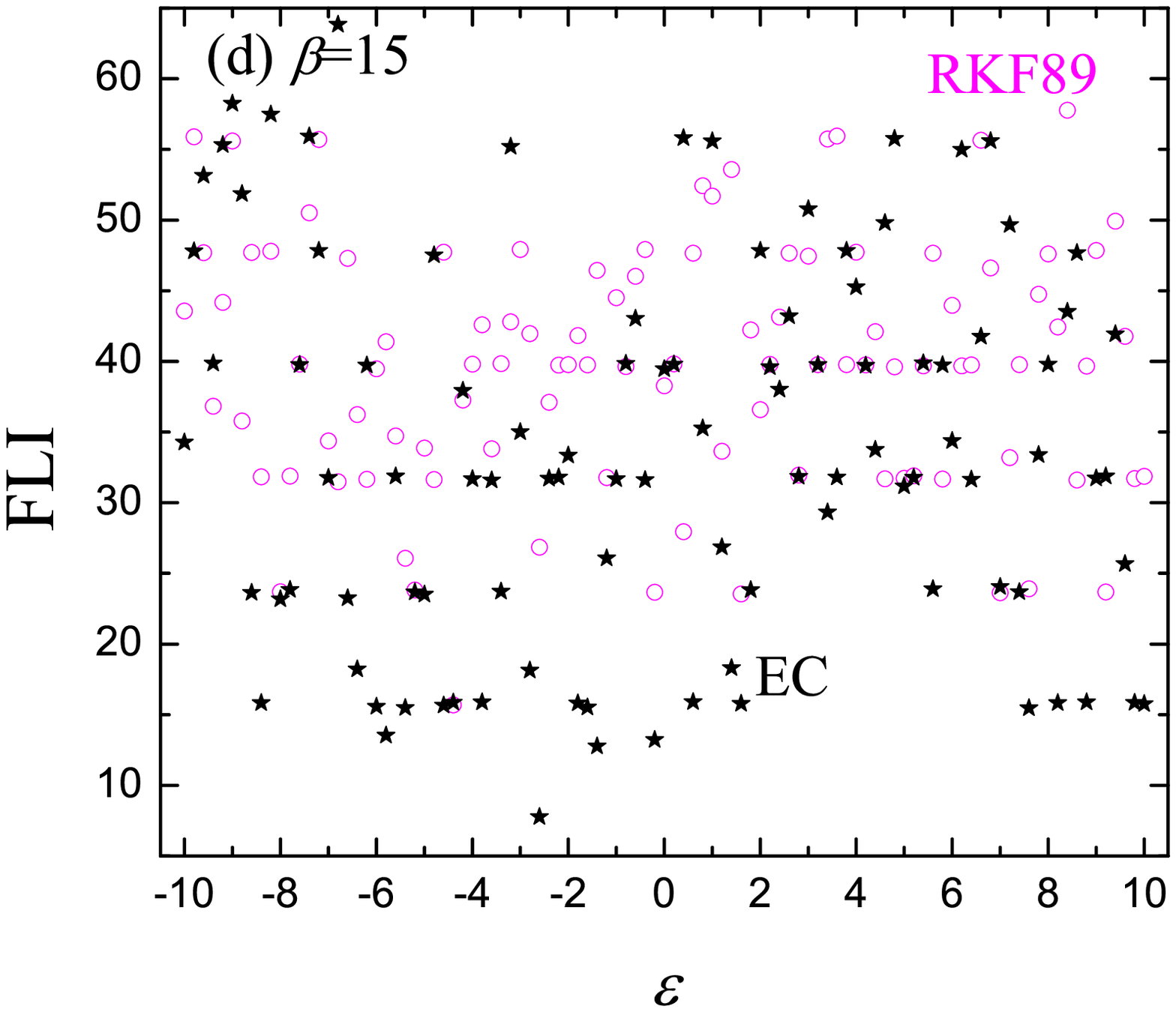}
\caption{As Figure 4, but relating to the dependence of FLIs on
parameters $\varepsilon_{m}=\varepsilon$.}} \label{fig5}
\end{figure*}

\begin{figure*}
\center{
\includegraphics[scale=0.4]{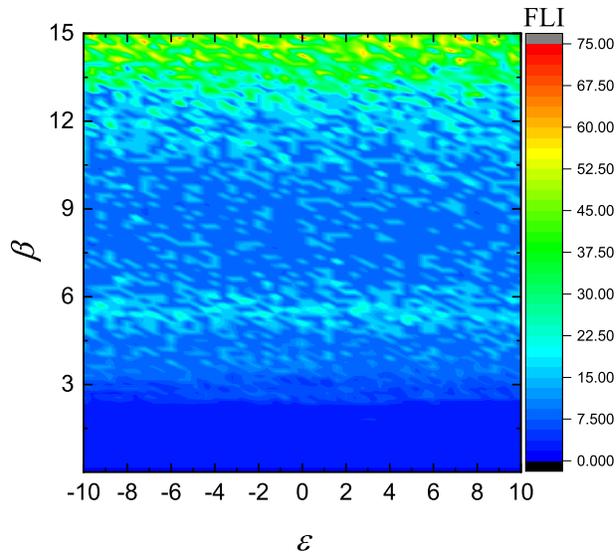}
\caption{Finding chaos by employing FLIs to scan a two-dimensional
space with parameters $\beta$ and $\varepsilon$ in the DDNLS
system.}} \label{fig6}
\end{figure*}

\begin{figure*}
\center{
\includegraphics[scale=0.3]{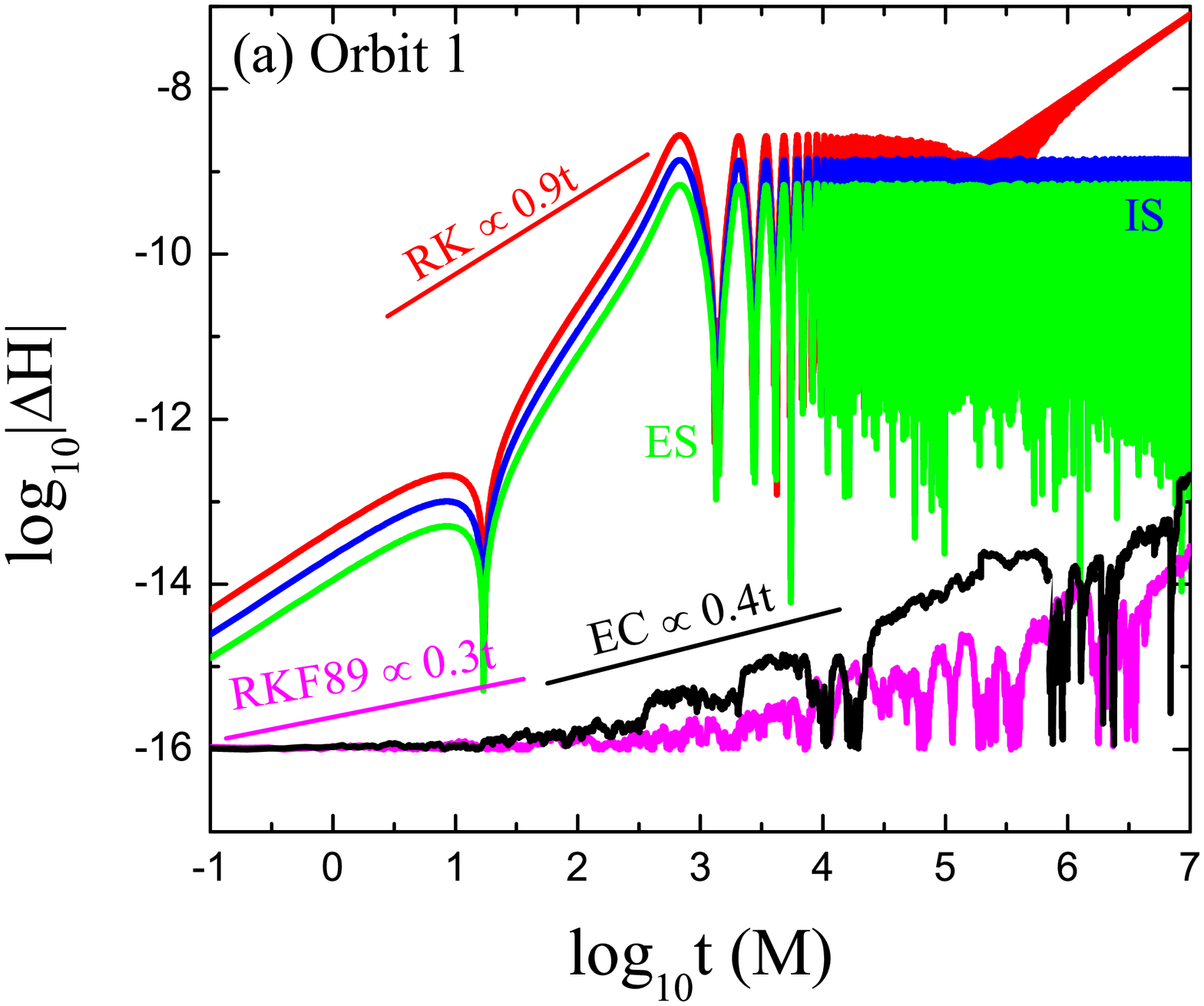}
\includegraphics[scale=0.3]{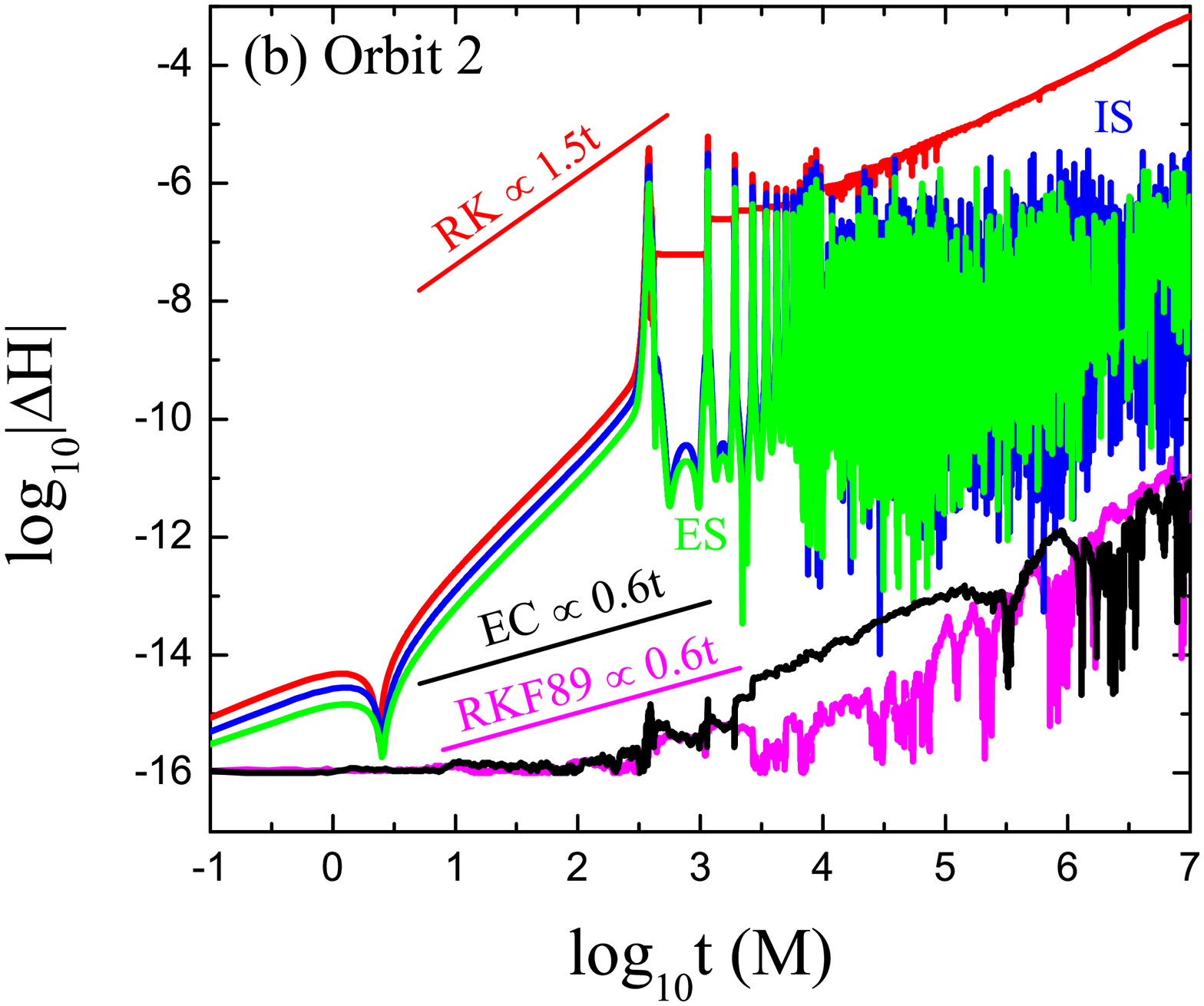}
\includegraphics[scale=0.3]{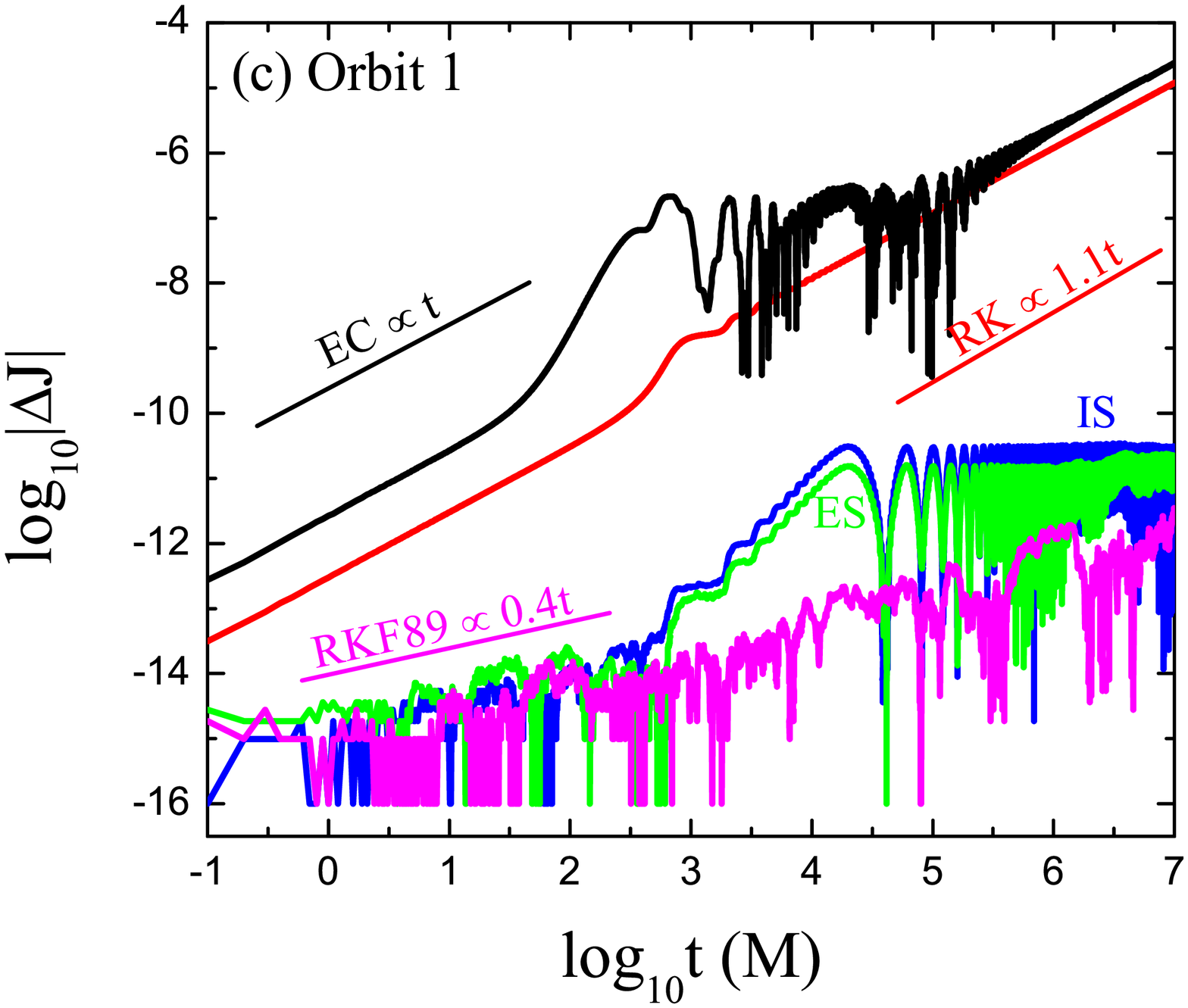}
\includegraphics[scale=0.3]{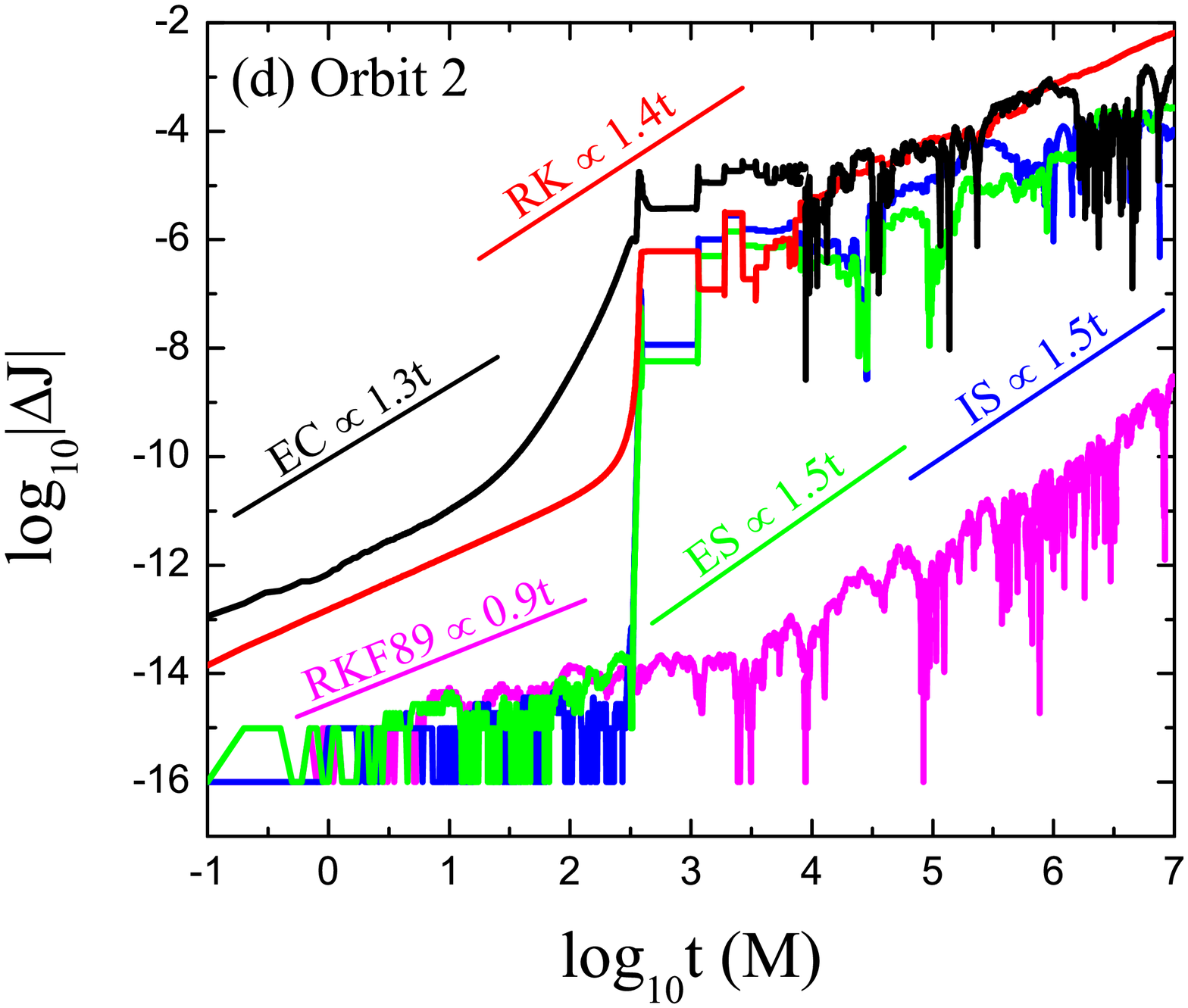}
\caption{Hamiltonian errors, $\Delta H$, for the four algorithms
solving Orbits 1 and 2 in the PN system of spinning compact
binaries. The time step is $h=0.1$, and the mass ratio is
$\gamma=1$. The two orbits have the initial conditions $x=40$,
$y=z=p_{x}=p_{z}=0$, $p_{y}=\sqrt{(1-e)/x}$,
$\theta_{1}=\theta_{2}=\pi/4$, and $\xi_{1}=\xi_{2}=0.1$. The
initial eccentricities are $e=0.0985$ for Orbit 1, and $e=0.7098$
for Orbit 2. EC and RKF89 exhibit virtually the same energy errors.
}} \label{fig7}
\end{figure*}

\begin{figure*}
\center{
\includegraphics[scale=0.25]{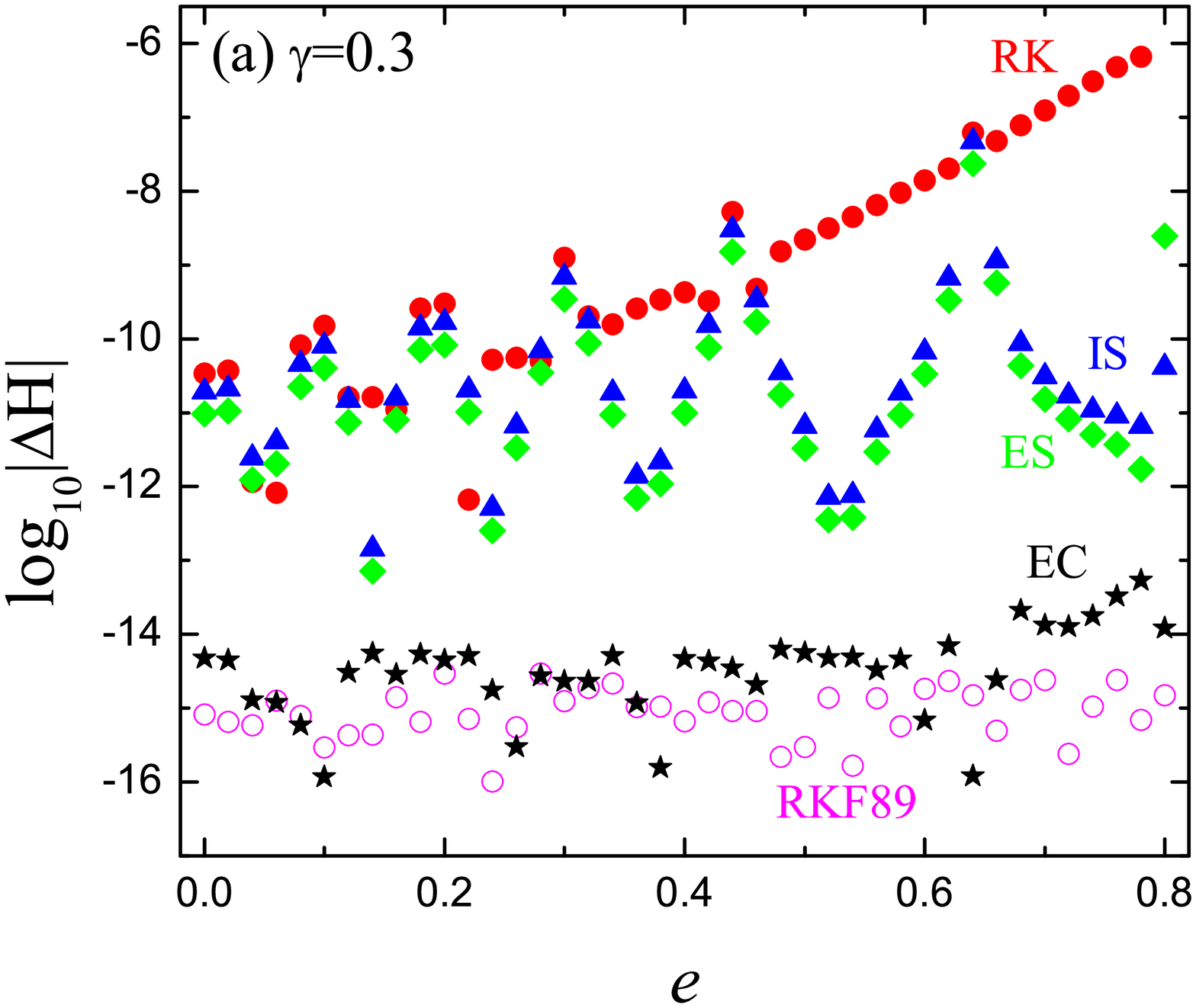}
\includegraphics[scale=0.25]{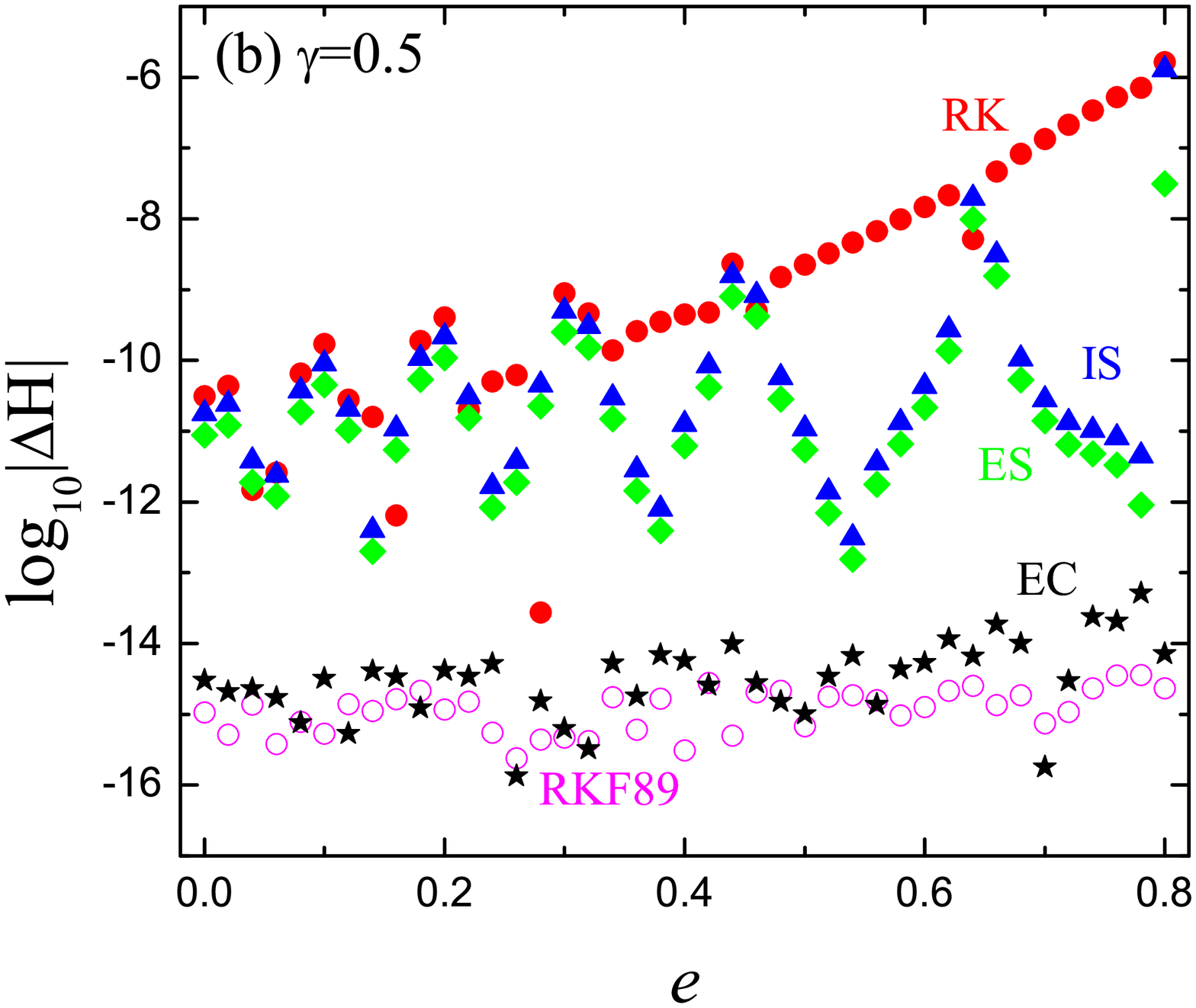}
\includegraphics[scale=0.25]{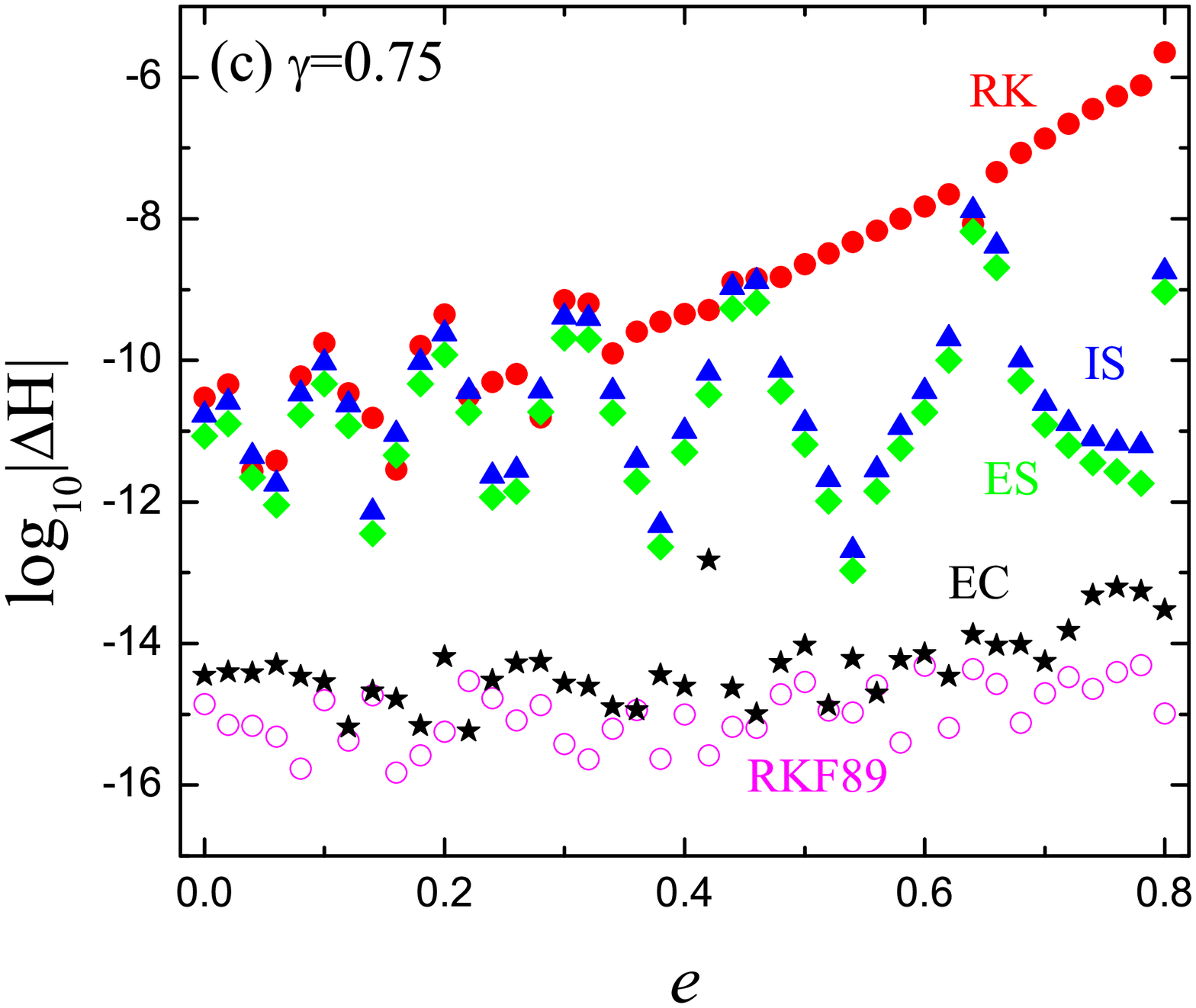}
\includegraphics[scale=0.25]{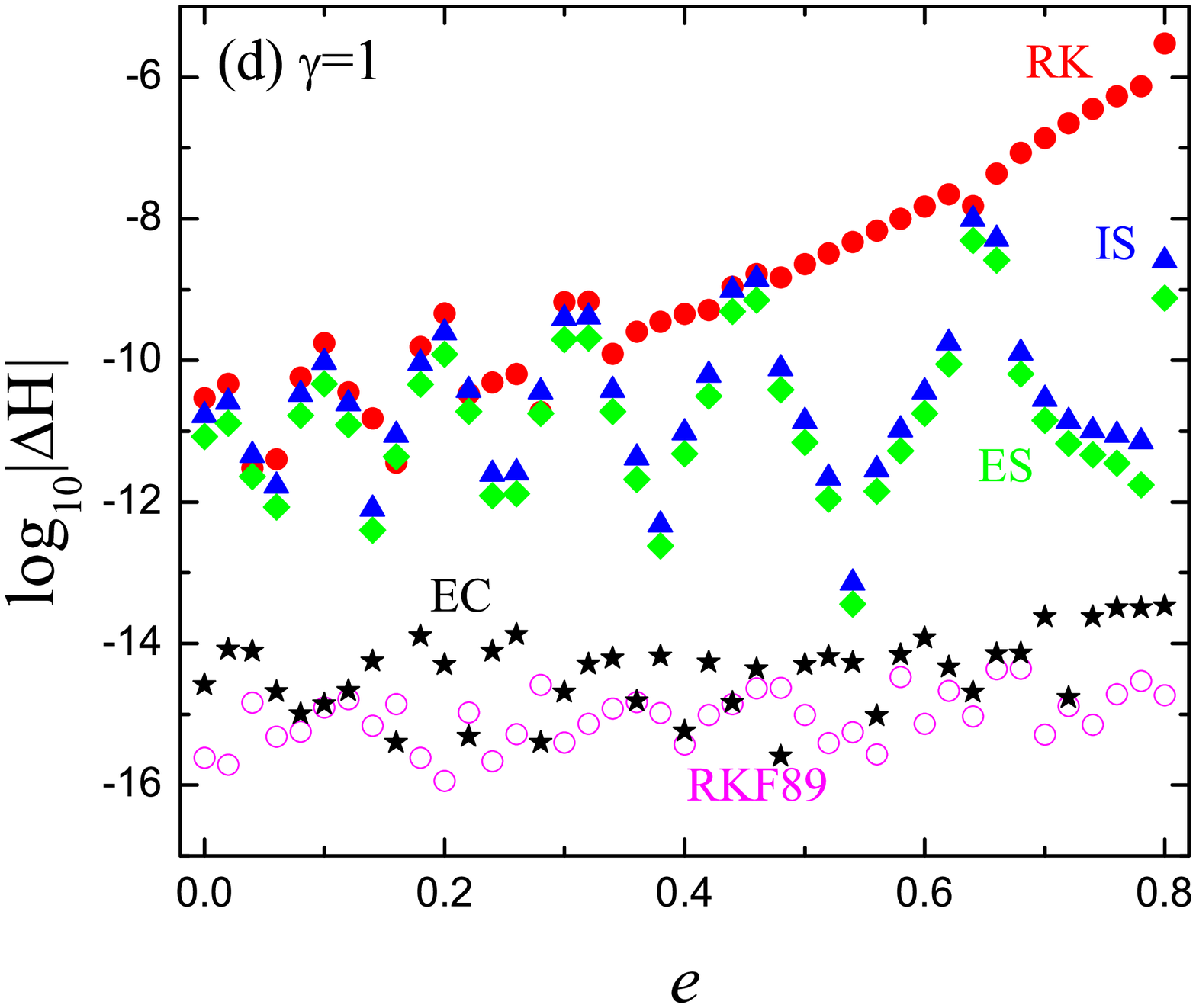}
\caption{Relations between Hamiltonian errors $\Delta H$ for the
five methods, with initial eccentricities, $e$. The same time step,
$h=0.1$, is used for EC, IS, ES, and RK, but different mass ratios
$\gamma$ appear in the four panels. Given a value of $e$, the
error is obtained after the integration time $t= 10^{5}$. The
results for EC are basically consistent with those for RKF89.}}
\label{fig8}
\end{figure*}

\begin{figure*}
\center{
\includegraphics[scale=0.4]{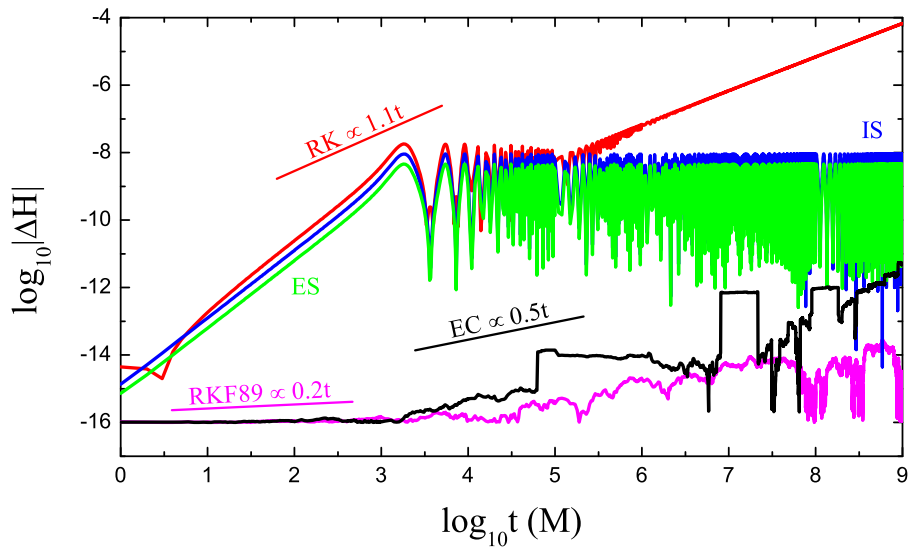}
\caption{Hamiltonian errors, $\Delta H$, for the five algorithms
solving Orbit 3 in the PN system of spinning compact binaries.
The time step is $h=1$, and the mass ratio is $\gamma=1$. This
orbit has the initial conditions $x=80$, $y=z=p_{x}=p_{z}=0$, and
$p_{y}=\sqrt{(1-e)/x}$, where $e=0.15$, and
$\theta_{1}=\theta_{2}=\xi_{1}=\xi_{2}=0$. After $10^9$
integration steps, IS and ES show no secular change in their energy
errors. EC, like RKF89, shows a secular change in terms of energy errors,
but exhibits smaller energy errors than either IS or ES.}} \label{fig9}
\end{figure*}

\begin{figure*}
\center{
\includegraphics[scale=0.25]{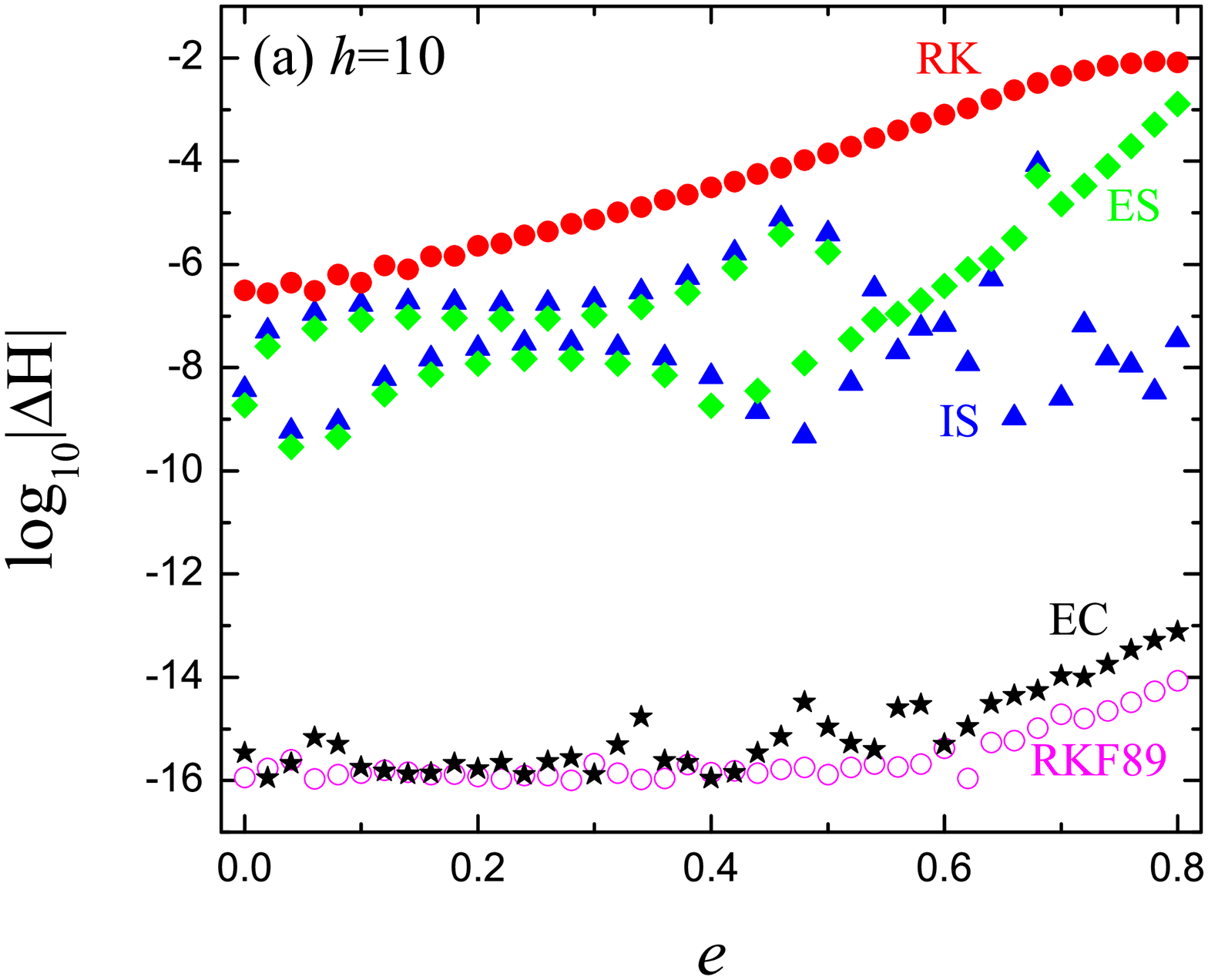}
\includegraphics[scale=0.25]{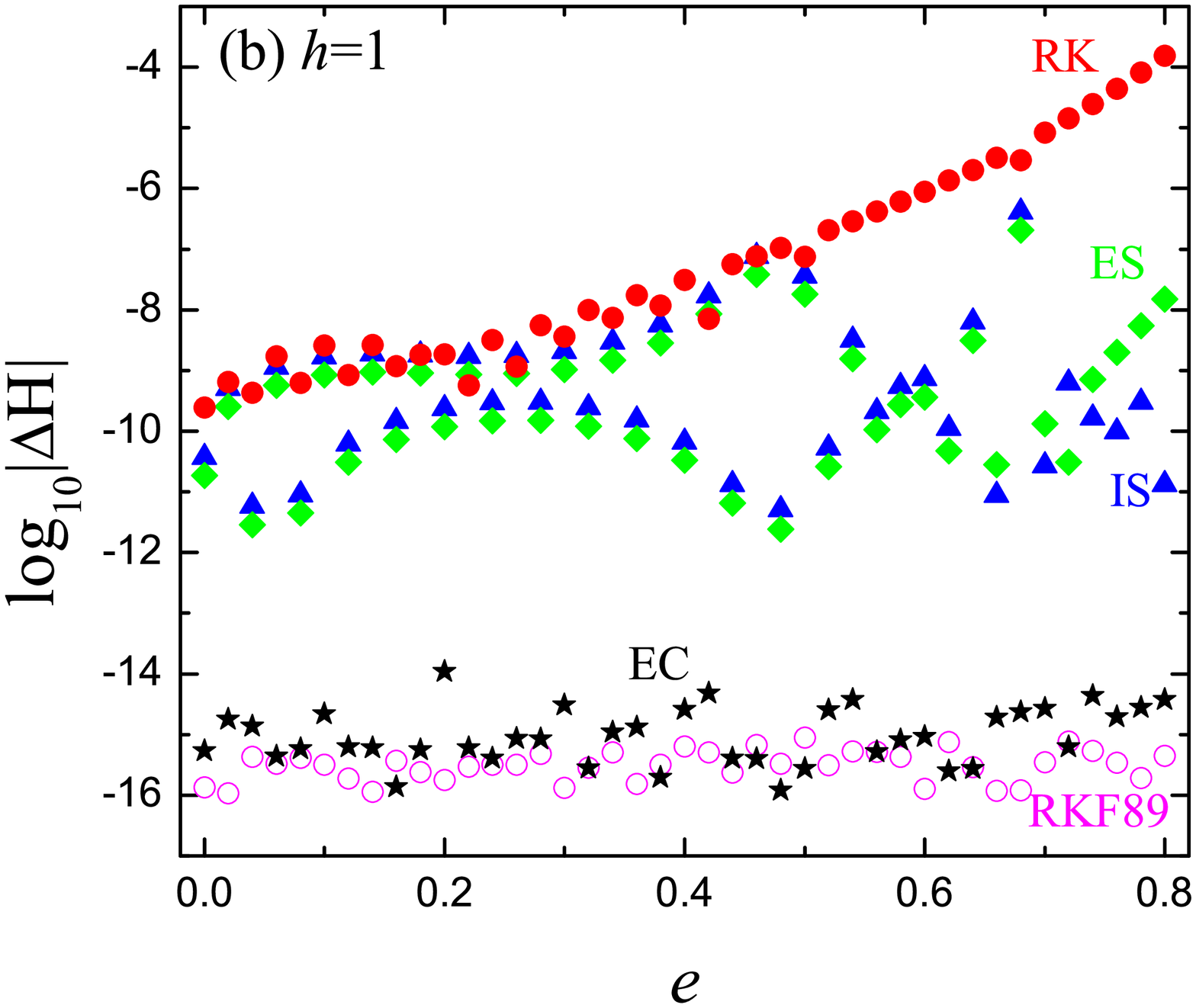}
\includegraphics[scale=0.25]{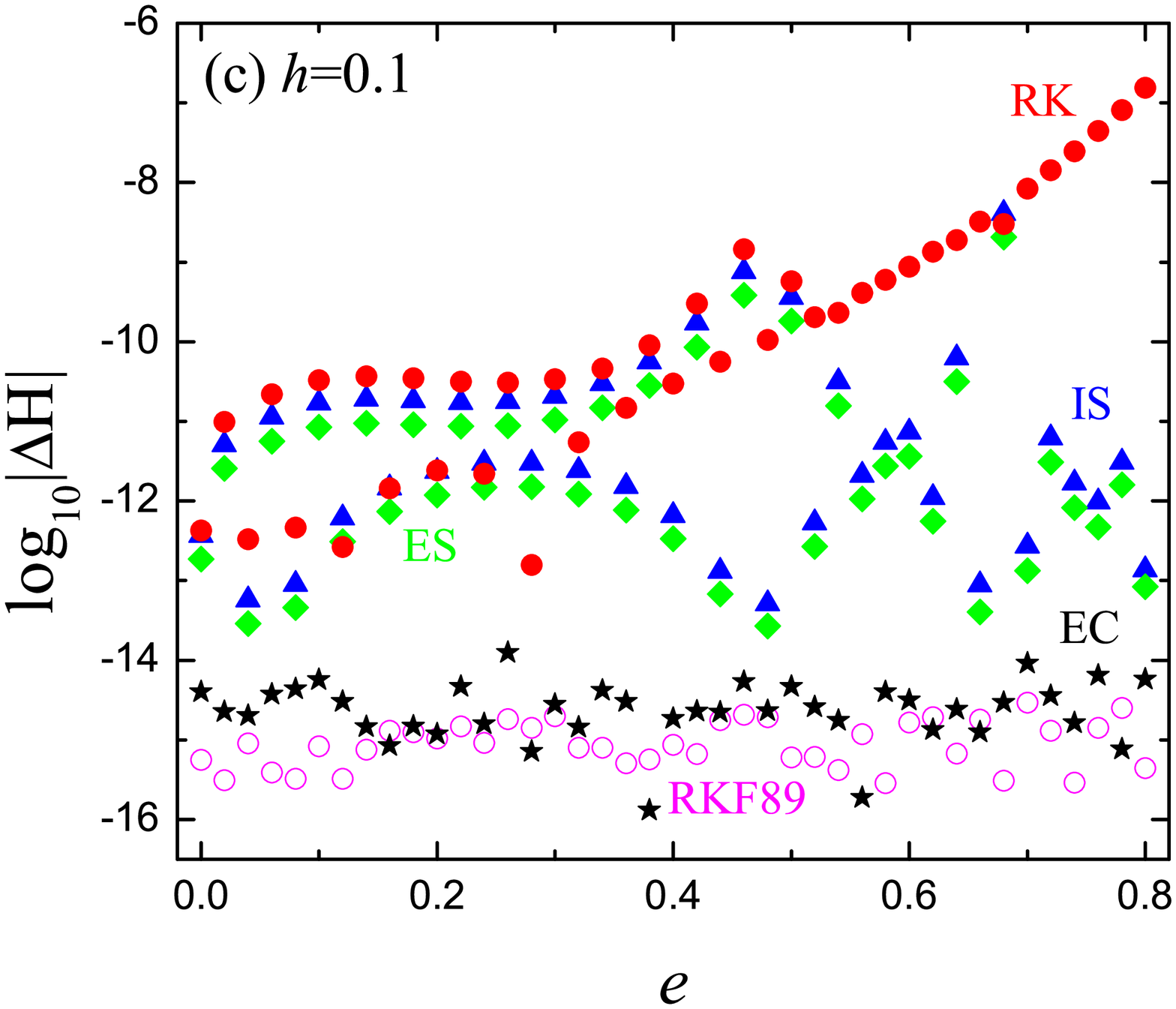}
\includegraphics[scale=0.25]{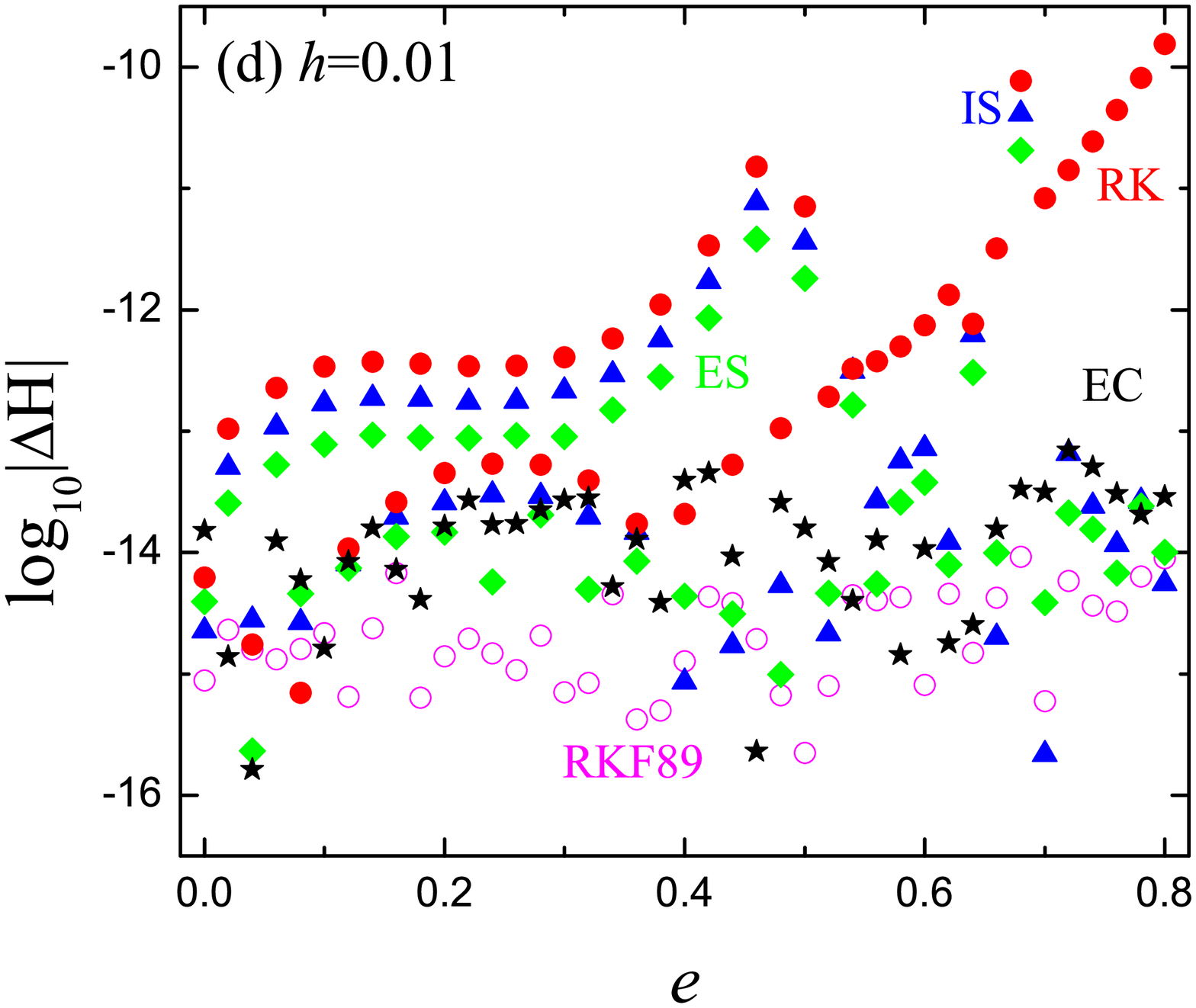}
\caption{Figure 8 continued. The same mass ratio, $\gamma=1$, is used,
but different time steps, $h$, are given. }} \label{fig10}
\end{figure*}

\begin{figure*}
\center{
\includegraphics[scale=0.4]{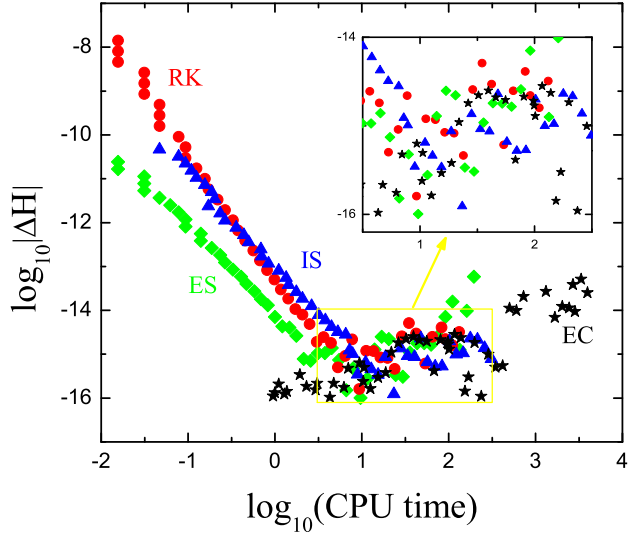}
\caption{As Figure 2, but with the initial conditions $x=150$,
$y=z=p_{x}=p_{z}=0$, $p_{y}=0.082745147$, and
$\theta_{1}=\theta_{2}=\xi_{1}=\xi_{2}=0$. The maximum error is
obtained after the integration time $t=10^{5}$. The time steps are
fixed for each algorithm, but the points correspond to different
time steps, $h=10/1.20679264^{k-1}$, where $k=1,2,\cdots,50$.
Although EC exhibits the poorest efficiency, its accuracy does not
depend on the choice of step size. }} \label{fig11}
\end{figure*}

\begin{figure*}
\center{
\includegraphics[scale=0.2]{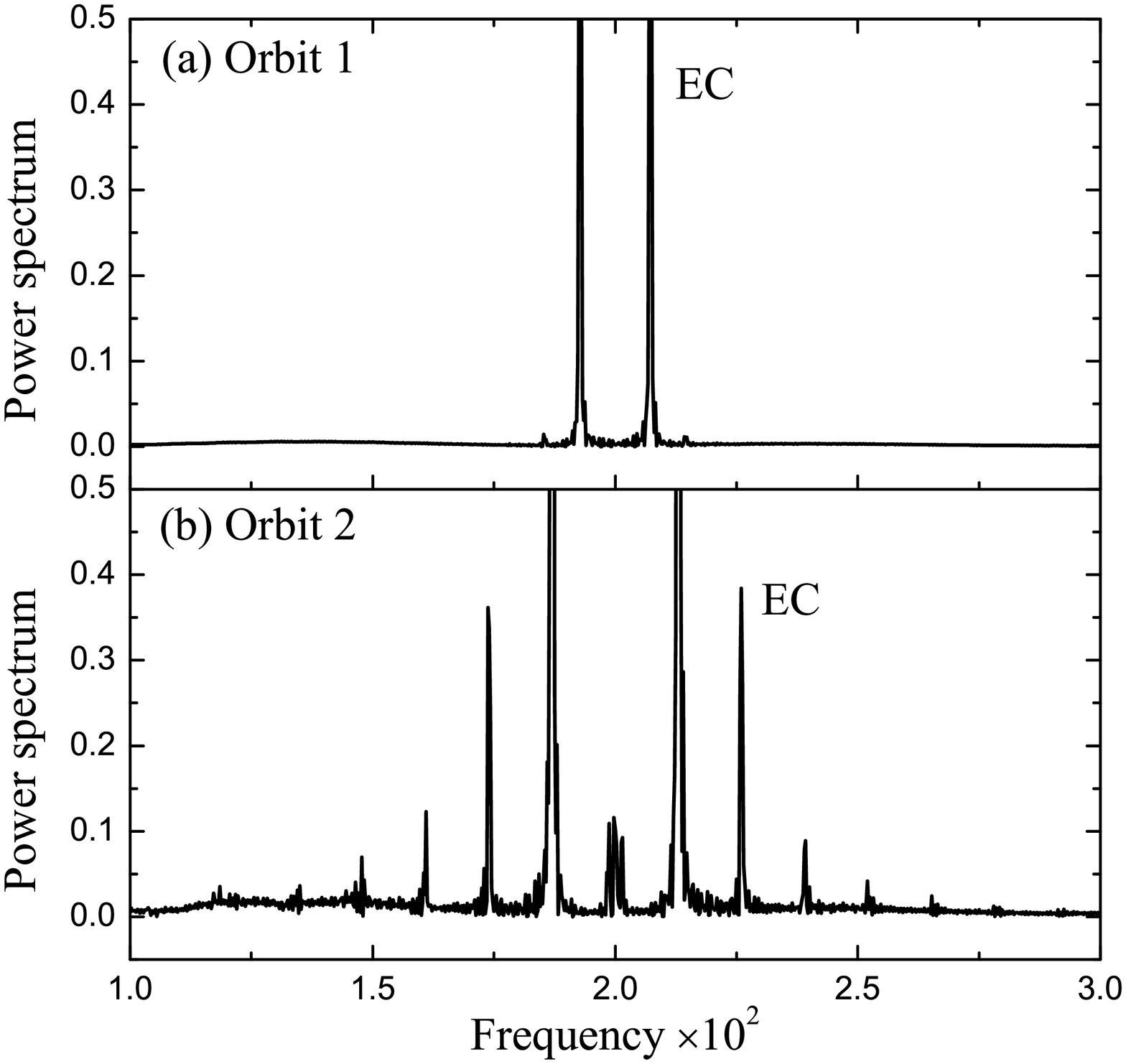}
\includegraphics[scale=0.2]{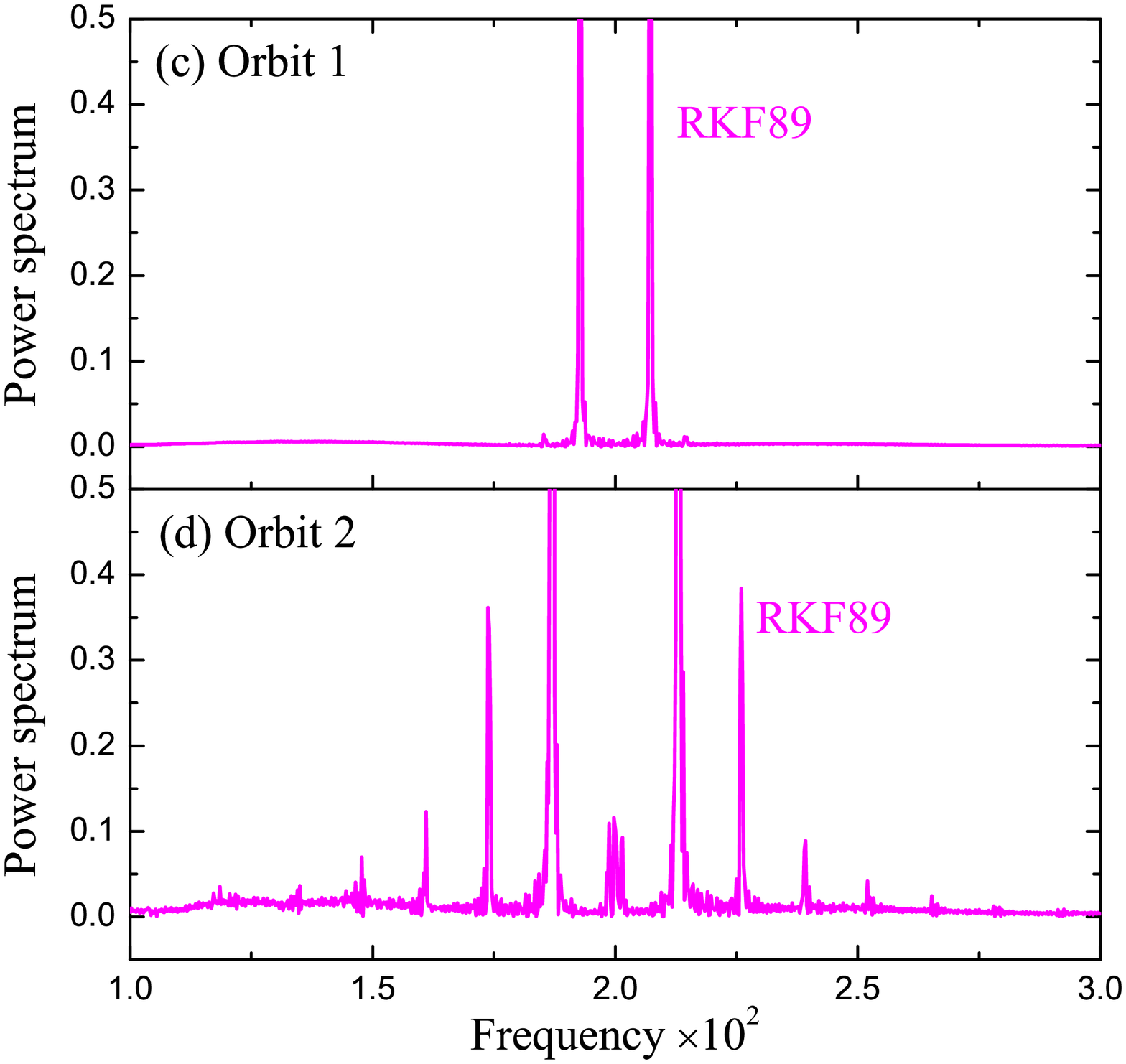}
\includegraphics[scale=0.2]{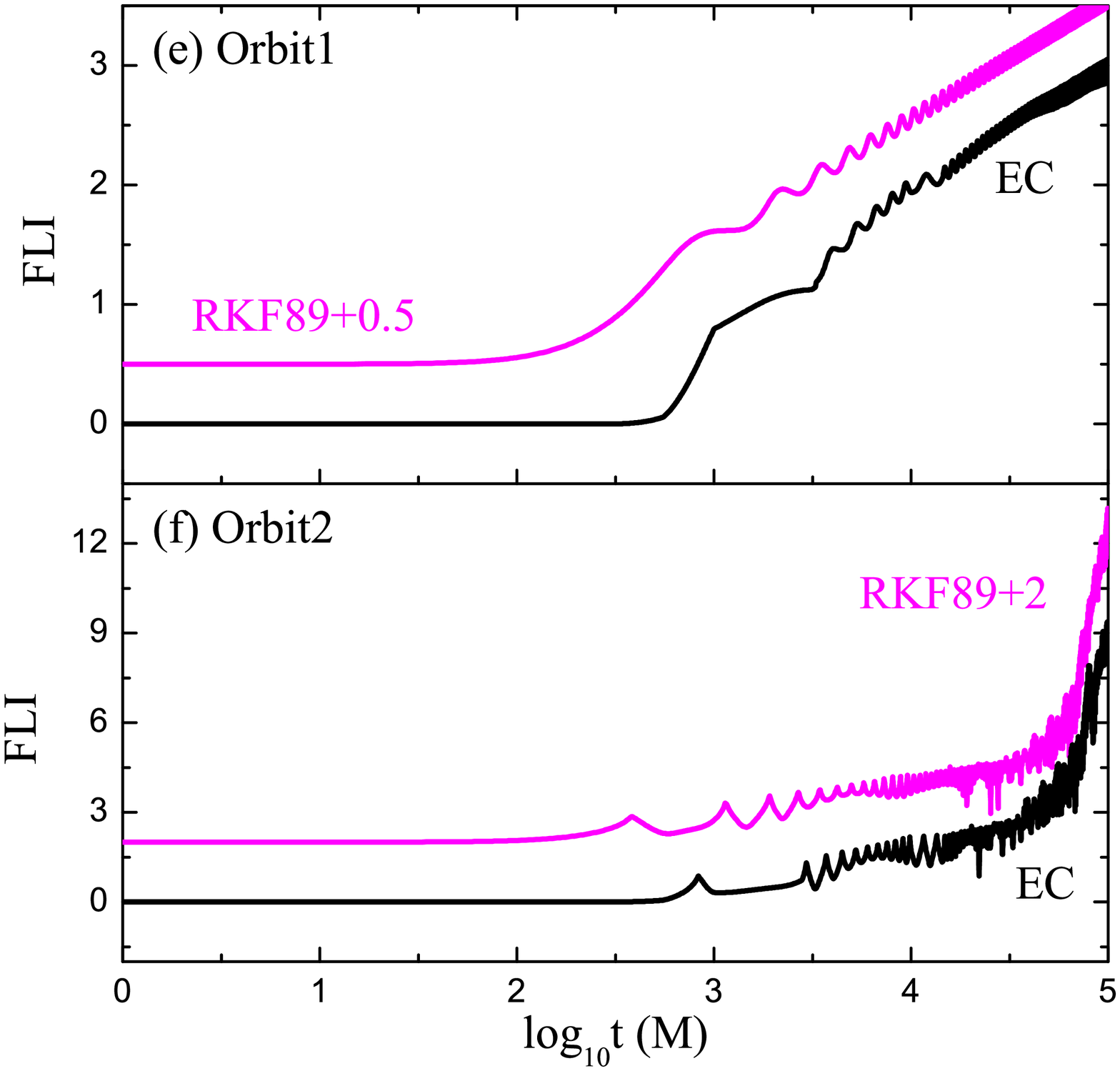}
\caption{Power spectra and FLIs of Orbits 1 and 2 in the PN
spinning binary system, obtained via the EC and RKF89 methods. Both
show that Orbit 1 is ordered, and Orbit 2 is chaotic and EC and RKF89
produce the same results. Note that EC and RKF89 approximately
coincide in panel (e).}} \label{fig12}
\end{figure*}

\begin{figure*}
\center{
\includegraphics[scale=0.3]{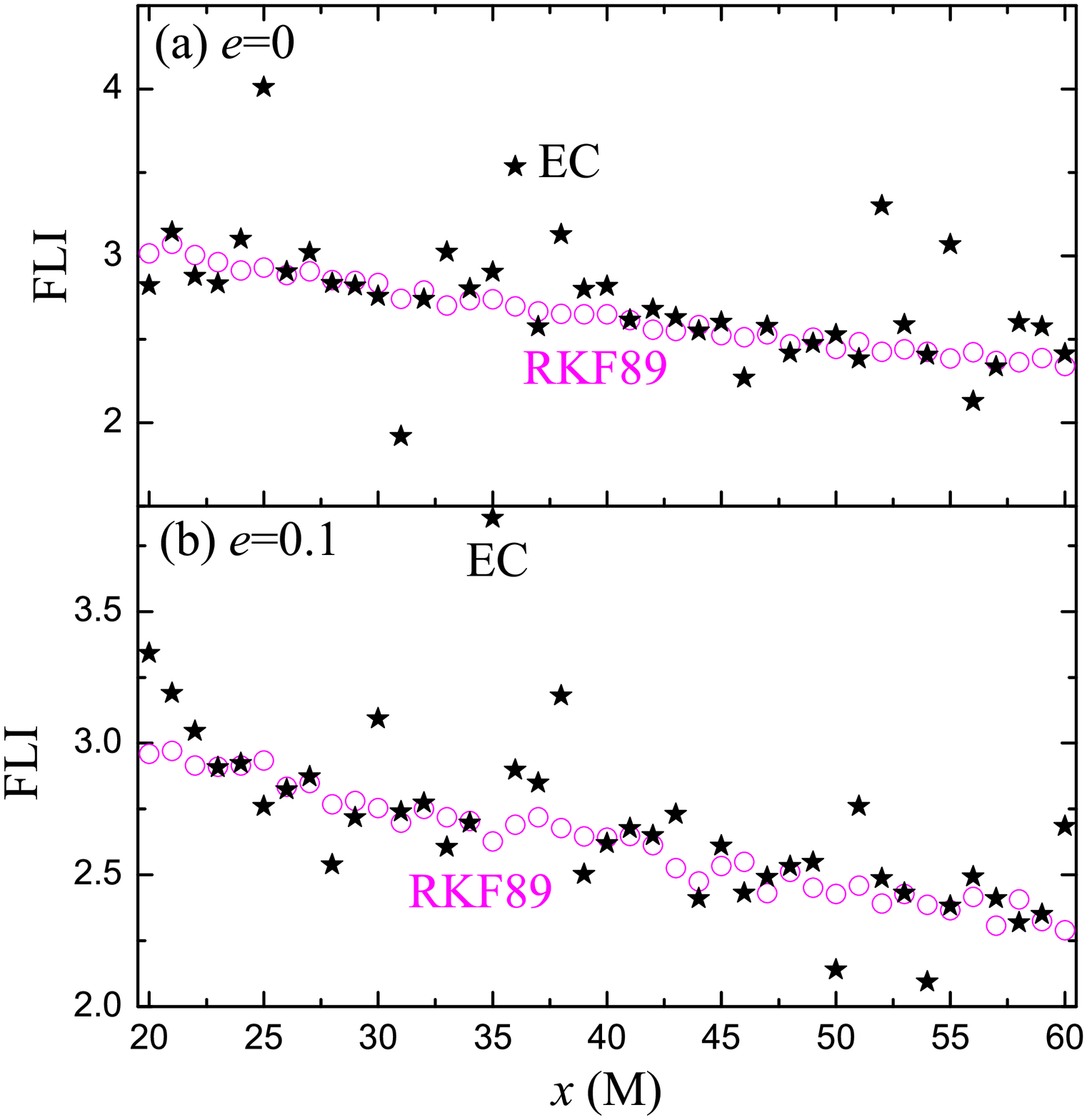}
\includegraphics[scale=0.3]{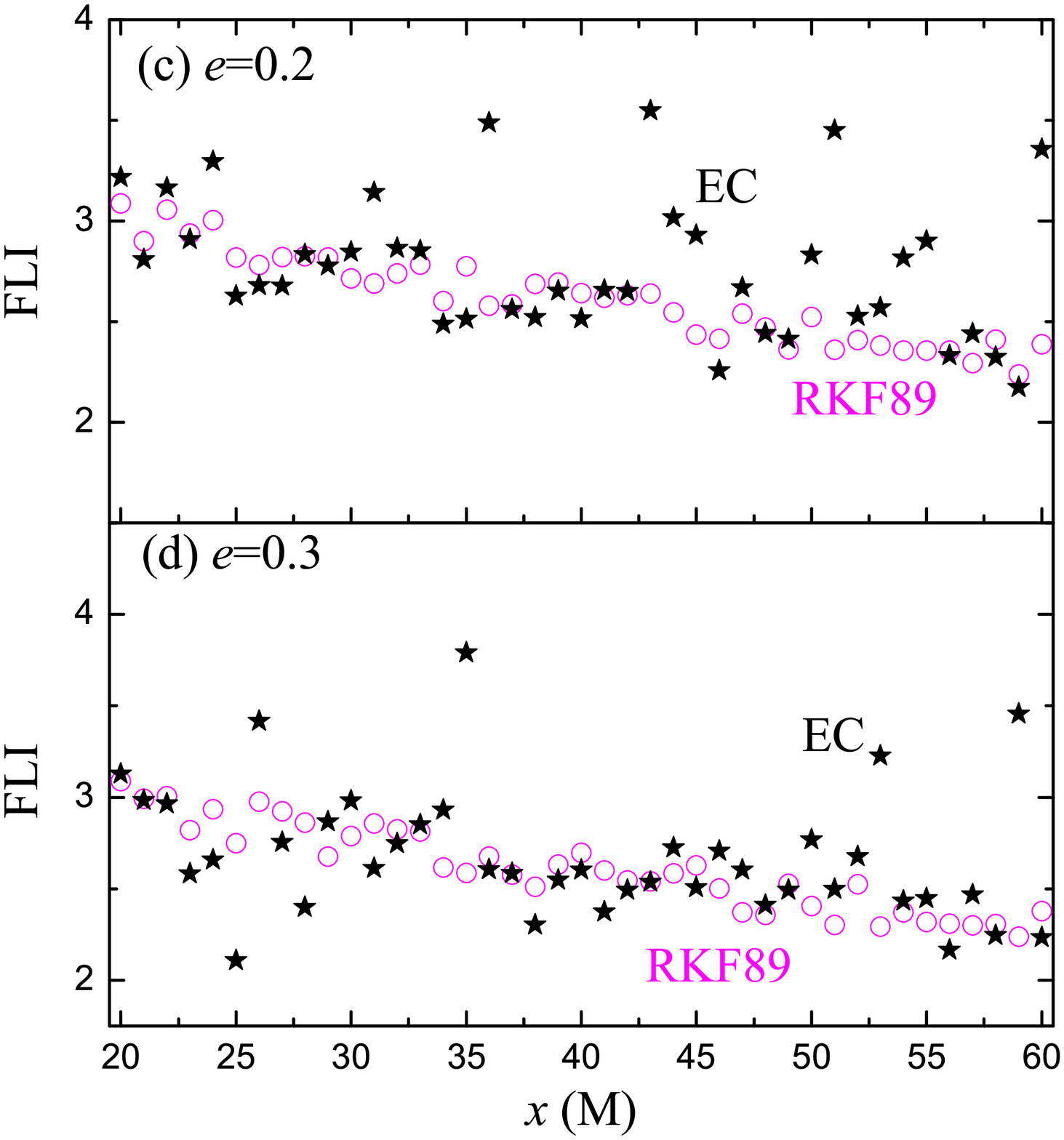}
\includegraphics[scale=0.3]{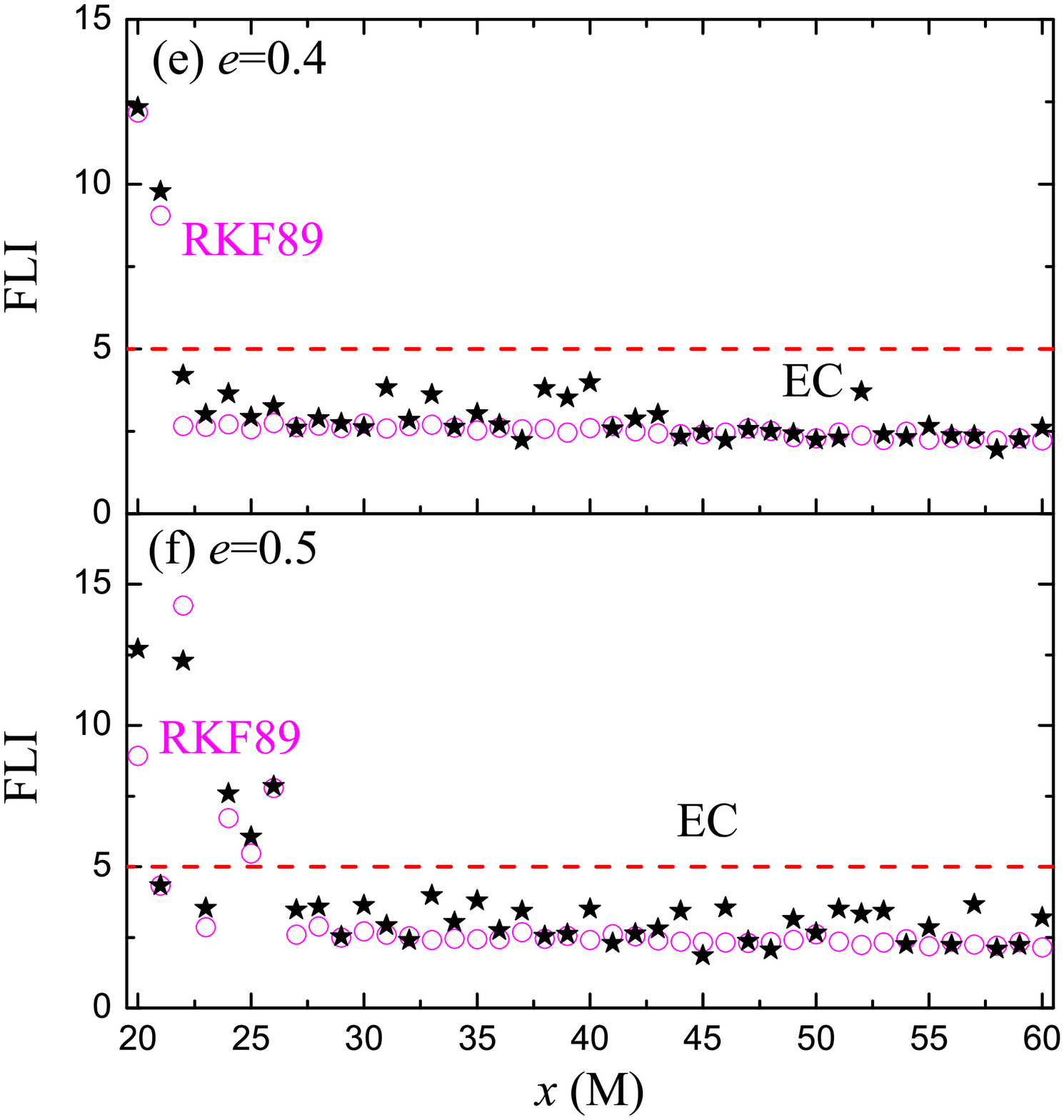}
\includegraphics[scale=0.3]{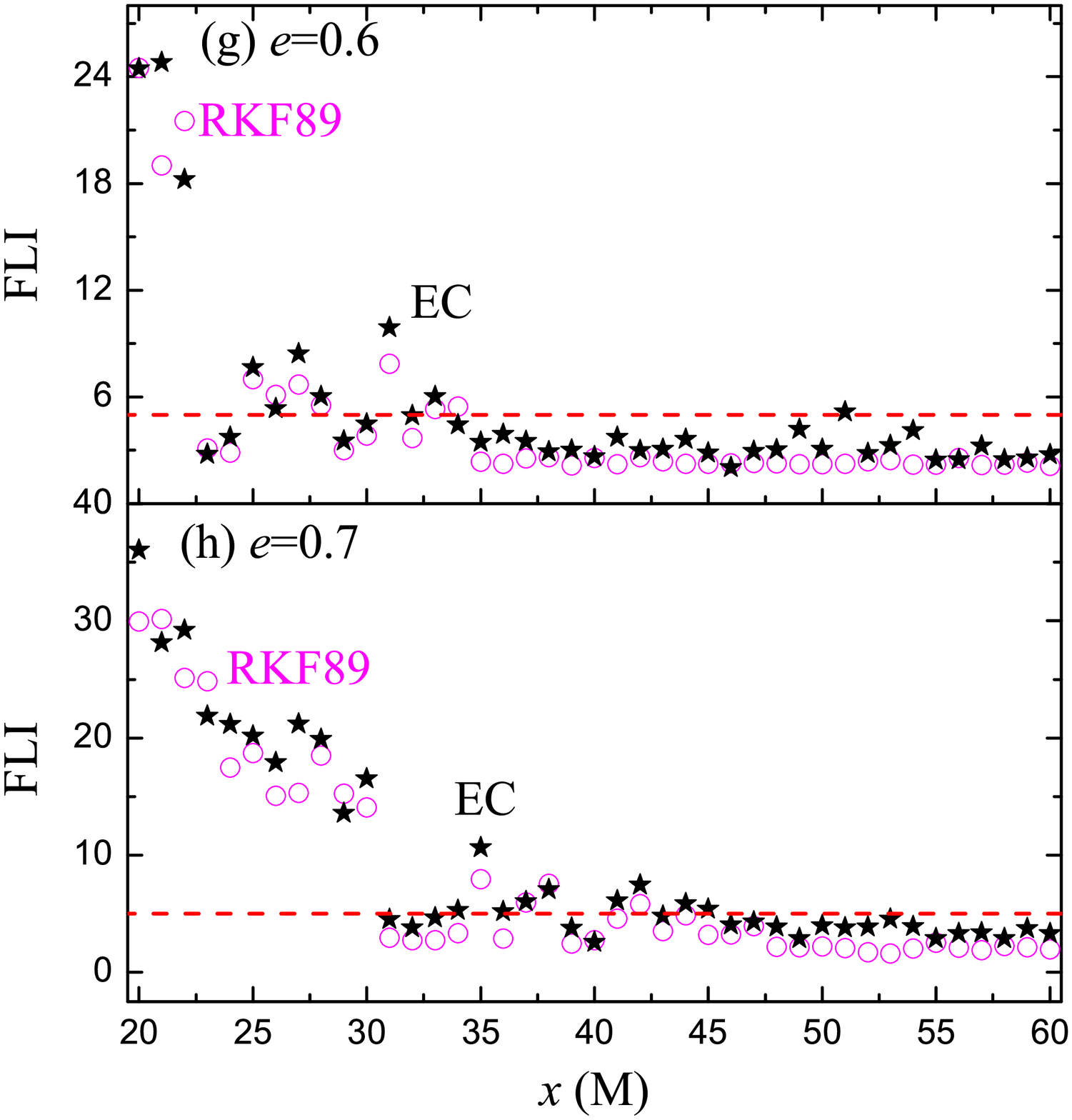}
\caption{Relations between FLIs and initial separations, $x$. Given
an initial separation $x$, the FLI is obtained after the
integration time  $t=3.5 \times 10^{4}$. Here, FLI=5 still represents the threshold between ordered and chaotic cases. Chaos readily occurs
for small initial separations and large initial eccentricities.
The results for EC are the same as those for RKF89.}} \label{fig13}
\end{figure*}

\begin{figure*}
\center{
\includegraphics[scale=0.4]{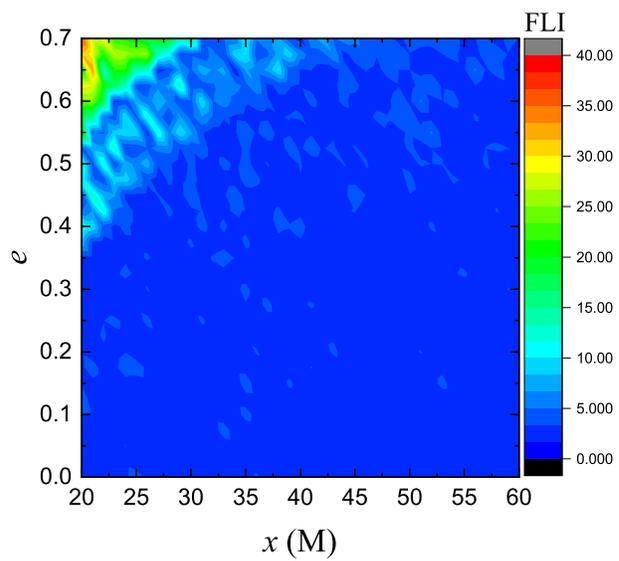}
\caption{{Finding chaos by employing FLIs to scan  a
two-dimensional space of initial eccentricity $e$, and initial
separation, $x$, for a PN spinning binary system. The mass ratio is
$\gamma=1$.}}} \label{fig14}
\end{figure*}

\end{document}